\documentclass{aastex}

\usepackage{graphicx}

\usepackage{emulateapj5}
\usepackage{times}

\makeatletter

\newenvironment{inlinefigure}{%
\def\@captype{figure}%
\noindent\begin{minipage}{0.999\linewidth}\begin{center}}
{\end{center}\end{minipage}\smallskip}
\makeatother

\usepackage{mathptmx}

\def\lax    {{_<\atop^{\sim}}}

\begin{document}

\slugcomment{accepted for publication in {\em The Astrophysical Journal}}

\title{The Extended Fe distribution in the intracluster medium 
and the implications regarding AGN Heating.}

\author{Laurence P. David \& Paul E.J. Nulsen}
\affil{Harvard-Smithsonian Center for Astrophysics, 60 Garden St.,
Cambridge, MA 02138;\\ david@cfa.harvard.edu}

\shorttitle{\emph Fe Distribution in Clusters}

\begin{abstract}
We present a systematic analysis of XMM-Newton observations of 8 cool-core
clusters of galaxies and determine the Fe distribution in the intracluster
medium relative to the stellar distribution in the central dominant galaxy (CDG).
Our analysis shows that the Fe is significantly more extended than the stellar mass
in the CDG in all of the clusters in our sample, with a slight trend of increasing
extent with increasing central cooling time. The excess Fe within the
central 100~kpc 
%%(i.e., above the mean Fe abundance at larger radii) 
in these
clusters can be produced by Type Ia supernovae from the CDG over the past 3-7~Gyr.
Since the excess Fe primarily originates from the CDG, it is a useful probe for
determining the motion of the gas and the mechanical energy deposited
by AGN outbursts over the past $\sim$~5~Gyr in the centers of clusters.
We explore two possible mechanisms for producing the greater extent of
the Fe relative to the stars in the CDG, including: bulk expansion of the gas and
turbulent diffusion of Fe.
Assuming the gas and Fe expand together,
we find that a total energy 
%%(thermal plus potential) 
of $10^{60} - 10^{61}$~erg~s$^{-1}$ must
have been deposited into the central 100~kpc of these clusters to
produce the presently observed Fe distributions.
Since the required enrichment time for the excess Fe is approximately 5~Gyr in these
clusters, this gives an average AGN mechanical power over this time of
$10^{43} - 10^{44}$~erg~s$^{-1}$.
The extended Fe distribution in cluster cores can also arise from
turbulent diffusion.  Assuming steady-state (i.e., the outward mass flux of
Fe across a given surface is equal to the mass injection rate of Fe within
that surface) we find that diffusion coefficients of $10^{29} - 10^{30}$~cm$^2$~s$^{-1}$
are required to maintain the presently observed Fe profiles.
We find that heating by both turbulent diffusion of entropy and dissipation
are important heating mechanisms in cluster cores.  In half of the clusters with
central cooling times greater than 1~Gyr, we find that heating by turbulent
diffusion of entropy alone can balance radiative losses.   In the remaining clusters,
some additional heating by turbulent dissipation, with turbulent velocities
of 150 - 300~km~s$^{-1}$, is required to balance radiative cooling.
We also find that the average Type Ia supernova fraction within the central 100~kpc
of these clusters is 0.53 (roughly twice the solar value) based on the
Si-to-Fe mass ratio.  This implies a total (Type Ia plus core collapse) supernova
heating rate less than 10\% of the bolometric X-ray luminosity within the centers of
clusters.
\end{abstract}


%% subject header.  Objects should be in the appropriate "individual"
%% headers (e.g. quasars: individual, stars: individual, etc.) with the
%% additional provision that the total number of headers, including each
%% individual object, not exceed six.  The \objectname{} macro, and its
%% alias \object{}, is used to mark each object.  The macro takes the object
%% name as its primary argument.  This name will appear in the paper
%% and serve as the link's anchor in the electronic edition if the name
%% is recognized by the data centers.  The macro also takes an optional
%% argument in parentheses in cases where the data center identification
%% differs from what is to be printed in the paper.



\section{Introduction}

There is compelling evidence from Chandra observations that heating
by a central AGN has a dramatic impact on the energetics of the hot gas
in early-type galaxies, groups, and clusters of galaxies
(McNamara \& Nulsen 2007; McNamara et al. 2000, Finoguenov \& Jones 2002;
Churazov et al. 2002;
Fabian et al. 2003, Blanton et al. 2003,
Mazzotta, Edge \& Markevitch 2003; Kraft et al. 2003;
Nulsen et al. 2005a, 2005b; Forman et al. 2005, Wise et al. 2007).
Heating from AGN driven shocks and buoyantly rising bubbles filled with radio
emitting plasma will also affect the distribution of the hot gas and heavy
elements in the central regions of clusters.
Einstein and Rosat observations showed that the hot gas in groups is more
extended (compared to the dark matter) than the gas in rich clusters
(David, Jones \& Forman 1995; Jones \& Forman 1999).  The variable extent of the
hot gas relative to the dark
matter breaks the self-similarity between groups and rich clusters.
The origin of this self-similarity breaking has usually been attributed to some
form of non-gravitational heating or ``pre-heating"
(David, Jones \& Forman 1995; Lloyd-Davies, Ponman \& Cannon 2000).
Recent estimates on the AGN mechanical heating rate, or ``cavity power", in
cool-core clusters (Birzan et al. 2004; Best et al. 2006; Rafferty et al. 2006;
Dunn \& Fabian 2008)
and the large radius of the shock in Hydra A (Nulsen et al. 2005a) suggest that much
of the pre-heating required to break the self-similarity of clusters could arise
during the formation and subsequent growth of supermassive black holes in the
host galaxies in clusters.

Observations by ASCA, {\it Beppo}SAX, XMM-Newton, and Chandra have shown
that clusters with cool cores have Fe abundance profiles that increase
toward the cluster center, while non cool-core clusters
(which have probably experienced a recent merger) have flatter Fe
abundance profiles (Finoguenov, David \& Ponman 2000; Fukazawa et al. 2000;
De Grandi et al. 2004; Tamura et al. 2004; Vikhlinin et al. 2005; Baldi et al. 2007).
Discussions about the history of chemical enrichment in clusters of galaxies
can be found in Renzini et al. (1993), Loewenstein (2006), and Matteucci (2007).
The Fe is a good tracer for determining the past motion of the gas in the
central region of clusters since it primarily originates from Type Ia
supernovae (SNe Ia) from the CDG (Finoguenov et al. 2000; Finoguenov et al. 2001;
Bohringer et al. 2004; De Grandi et al. 2004).  Most studies have found that the
Fe is more extended than the light of the CDG (David et al. 2001;
Rebusco et al. 2005, 2006).
%%however, Sanders \& Fabian (2006) claimed that 
%%the Fe and light of the CDG
%%in the Centaurus cluster have similar distributions.
The greater extent of the Fe compared to the light of the CDG is likely the result
of AGN mechanical heating.  Tornatore et al. (2004) presented a series
of numerical simulations that tracked the chemical enrichment of the gas in
clusters and concluded that without any AGN feedback, the resulting Fe abundance
profiles are more centrally concentrated than those observed.

We present in this paper a systematic analysis of 8 relaxed, cool-core clusters
with measured optical surface brightness profiles for their CDG
and make a direct comparison between the

\begin{table*}[t]
\begin{center}
\caption{General Cluster Properties}
\begin{tabular}{lccccc}
\hline
Name & z & kT & $L_R(r<100 \rm{kpc})$ & $L_{bol}(r<100 \rm{kpc})$ & $t_c(r=10 \rm{kpc})$ \\
     & & (keV) & $(L_{\odot R})$ & (erg~s$^{-1}$) & (yr) \\
          \hline\hline
A262  & 0.0163 & 2.04 & $1.67 \times 10^{11}$ & $1.96 \times 10^{43}$ & $7.5 \times 10^{8}$ \\
A496  & 0.0329 & 3.23 & $4.44 \times 10^{11}$ & $1.29 \times 10^{44}$ & $6.0 \times 10^{8}$ \\
A2199 & 0.0301 & 3.77 & $4.12 \times 10^{11}$ & $1.52 \times 10^{44}$ & $6.8 \times 10^{8}$ \\
A2589 & 0.0414 & 3.39 & $2.81 \times 10^{11}$ & $4.85 \times 10^{43}$ & $2.3 \times 10^{9}$ \\
A3558 & 0.048  & 5.08 & $7.47 \times 10^{11}$ & $1.03 \times 10^{44}$ & $1.9 \times 10^{9}$ \\
A3571 & 0.0391 & 6.39 & $8.50 \times 10^{11}$ & $1.85 \times 10^{44}$ & $1.8 \times 10^{9}$ \\
A3581 & 0.023  & 1.62 & $1.72 \times 10^{11}$ & $2.78 \times 10^{43}$ & $6.6 \times 10^{8}$ \\
A4059 & 0.0475 & 3.47 & $5.13 \times 10^{11}$ & $9.31 \times 10^{43}$ & $1.1 \times 10^{9}$ \\
\hline
\end{tabular}

\noindent
Notes: Cluster name, redshift, emission-weighted temperature within 100~kpc,
R-band luminosity of the central dominant galaxy within 100~kpc,
bolometric X-ray luminosity within 100~kpc and isobaric radiative cooling time
at 10~kpc from the cluster center.
\end{center}
\end{table*}

\noindent
Fe and stellar mass
density distributions.  We find that the total Fe and the central Fe excess
are significantly more extended than the stars in the CDG
in all of the clusters in our sample. We investigate two possible origins for
the greater extent of the Fe relative to the stars: 
1) bulk expansion of the gas due to AGN heating
and 2) outward diffusion of Fe due to turbulent gas motions.

This paper is organized in the following manner. In $\S$~2, we present
our cluster sample and in $\S$~3 we discuss the details of our XMM-Newton
data reduction.  The total Fe mass, Si mass, SNe Ia fraction,
total supernova heating rate and the distribution of the excess Fe
relative to the stars in the CDG are presented in section $\S$~4. In
$\S$~5, we explore potential mechanisms for producing the
more extended Fe distribution, including: bulk expansion
of the gas due to AGN heating and turbulent diffusion of Fe.
The implications of our findings regarding the AGN cooling flow
feedback mechanism are discussed in $\S$~6.

\section{Cluster Sample}

We searched the XMM-Newton archive for deep observations of clusters
that have CDGs with optical
surface brightness profiles measured by Graham et al. (1996).
We excluded clusters with significant substructure and redshifts greater
than $z=0.05$ to ensure adequate spatial resolution of the Fe distribution.
Our final sample of 8 clusters is listed in Table 1.
We use XMM-Newton data in our study since good photon statistics are required
to accurately determine the Fe distribution.
The Fe abundance in these clusters typically decreases by a factor
of 2 over scales of $3-5^{\prime}$, so the $30^{\prime\prime}$
resolution of the XMM-Newton EPIC cameras for spectroscopic imaging is
sufficient for determining the Fe distribution in these clusters.
The photon statistics in most of the Chandra observations of these clusters
are insufficient to constrain the Fe distribution on smaller angular scales.

Graham et al. (1996) fit both Sersic and de Vaucouleurs profiles
to the R-band images of the CDGs in our sample and found that
most of the CDGs are better modeled with Sersic profiles, except
for the CDGs in A262 and A3581, which are adequately fit with de
Vaucouleurs profiles.  By fitting the 2-D images of a sample
of 24 higher redshift CDGs, Gonzalez, Zabludoff \& Zaritsky (2005) found that
the sum of two de Vaucouleurs profiles is a better representation
of the surface brightness profile compared to a
single Sersic model.  However, their Fig. 3 shows that
the azimuthally averaged surface brightness profiles are
equally well represented by either a Sersic model or the sum of
two de Vaucouleurs models.
Thus, the use of a
Sersic model for computing azimuthally averaged quantities should be sufficient.
Graham et al. list the effective radii for the CDGs in kpc based on the
distance indicator in Lauer \& Postman (1994).  We adjust these radii using
$\rm H_0$=70~km~s$^{-1}$~Mpc$^{-1}$, $\Omega_{M}$=0.3
and $\Omega_{\Lambda}$=0.7, which is the standard cosmology
adopted throughout this paper.  We correct the surface brightness at the
effective radius listed in Graham et al. for extinction and
apply a K-correction.  The extinction values are obtained from Schlegel et al.
(1998) and the K-correction,
appropriate for an early-type galaxy, is obtained from Postman \& Lauer (1995).
The stellar mass density profiles for the CDGs are then derived from the
Sersic profiles for all the CDGs except those in A262 and A3581, for which
we use the de Vaucouleurs profiles, using Abel's formula (Binney \& Tremaine 1987)
and $M_*/L_R = 3.5~M_{\odot}/L_{\odot R}$ (which is based on 
the average optical colors of the CDGs in our sample and the stellar synthesis 
models of Bell \& de Jong 2001).

\section{XMM-Newton Data Reduction}

All archived XMM-Newton data were reprocessed using the emchain and epchain
tasks in SASS v7.0 with the standard flag settings.  To screen for
background flares, we extracted light curves from each detector
in the 8-12~keV band within an annulus $10-12^{\prime}$ from the cluster
centers.  Time intervals with 8-12~keV count rates more the $3 \sigma$
above the mean count rate were excised from further analysis.
Background data sets were generated using the procedures outlined
in Arnaud et al. (2002) using the
blank field background data sets from Read \& Ponman (2003) and the
filter-closed charged particle background data sets from Marty et al. (2003).

After excising the emission from all detected point sources,
a background subtracted, exposure corrected 0.5-7.0~keV
surface brightness profile was generated for each cluster by
combining the data from all 3 EPIC cameras using the optical centroid of
the CDG as the origin.
Based on the cumulative number of net 0.5-7.0~keV counts, we generated
three sets of concentric annular regions that enclose at least 10000, 20000
and 40000 net counts with a minimum width for each annulus of $30^{\prime\prime}$.
For each cluster, source spectra were extracted from all 3 EPIC detectors
in the three sets of annular regions along with background
spectra from the same regions in the background data sets.
Photon weighted response and effective area files
were generated for 

\begin{inlinefigure}
\center{\includegraphics*[width=0.90\linewidth,bb=10 142 570 700,clip]{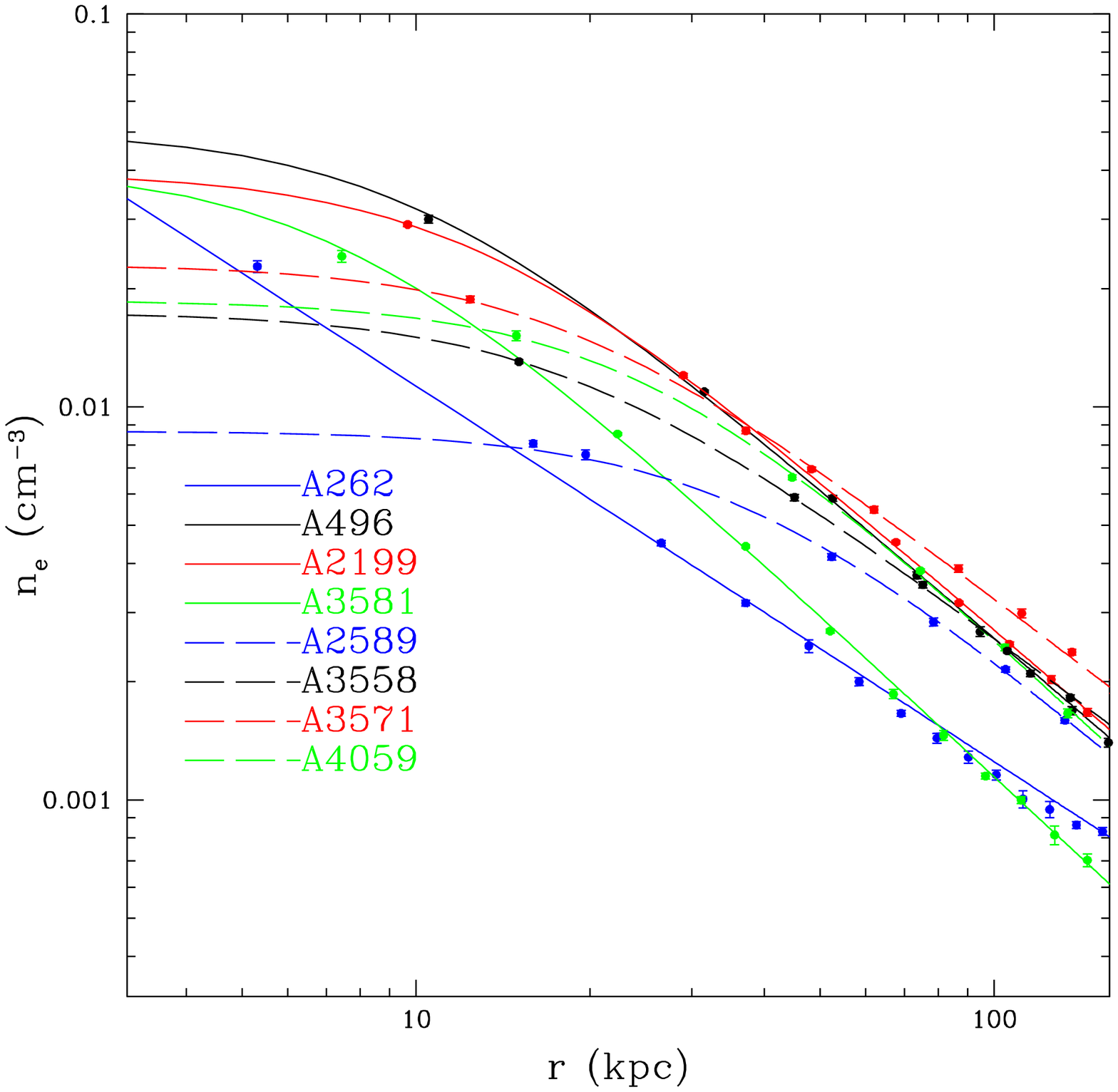}}
\caption{Electron number density vs. radius for each cluster in our sample.
Clusters with cooling times at 10~kpc less than 1~Gyr are
shown with solid lines and clusters with longer cooling times
are shown with dashed lines.}
\end{inlinefigure}

\bigskip
\noindent
each spectrum using the SAS tasks rmfgen and arfgen.
The three sets of spectra for each cluster were then deprojected assuming
spherical symmetry.  The spectra were fit to a single temperature
vapec model in the 0.8-8.0~keV bandpass
with the hydrogen column density frozen at the galactic value.
The abundances of all $\alpha$ processed elements along with the Fe abundance
and temperature were treated as free parameters.  The Ni abundance was linked to
the Fe abundance.  All abundances are measured relative to the solar values in
Grevesse \& Sauval (1998).
The gas density profile for each cluster is determined from the deprojected
emission measures of the set of spectra with at least 10,000 net counts.
The temperature and Fe abundance profiles are based on the deprojected values
from the set of spectra with at least 20,000 net counts.  The Si abundance profiles
are derived from the deprojected values from the set of spectra with at
least 40,000 net counts.  We also extracted a single spectrum from the
inner 100~kpc of each cluster to determine the emission-weighted temperature
and bolometric X-ray luminosity within this region (see Table 1).

\section{The Total and Excess Fe Mass Within the Central 100~kpc}

The general, azimuthally-averaged de-projected properties (density, temperature,
entropy, and Fe abundance profiles) of the clusters in our sample are shown in Figs. 1-4.
%%For clarity of presentation, we plot the cubic spline fit to the
%%deprojected quantities.  The wiggles in the curves simply reflect the
%%uncertainties in the best-fit values. 
We divide the clusters into two sub-samples based on the cooling
time of the hot gas at a radial distance of 10~kpc from
the CDG (see Table 1).  Clusters with cooling times less than 1~Gyr
are plotted with solid lines while clusters
with longer cooling times are plotted with dashed lines.
The deprojected density profiles were fitted to a single $\beta$ model and
the best-fit models are shown in Fig. 1. The deprojected temperature 
profiles were fitted with the analytic function used by Allen, 
Schmidt \& Fabian (2001) and the resulting profiles are shown in Fig. 2.  
For the Fe abundance profiles shown in Fig. 3, we fitted the data to a single $\beta$ model
plus a constant.  These figures show that the general cluster properties
can be characterized by their central cooling 

\begin{inlinefigure}
\center{\includegraphics*[width=0.90\linewidth,bb=10 142 570 700,clip]{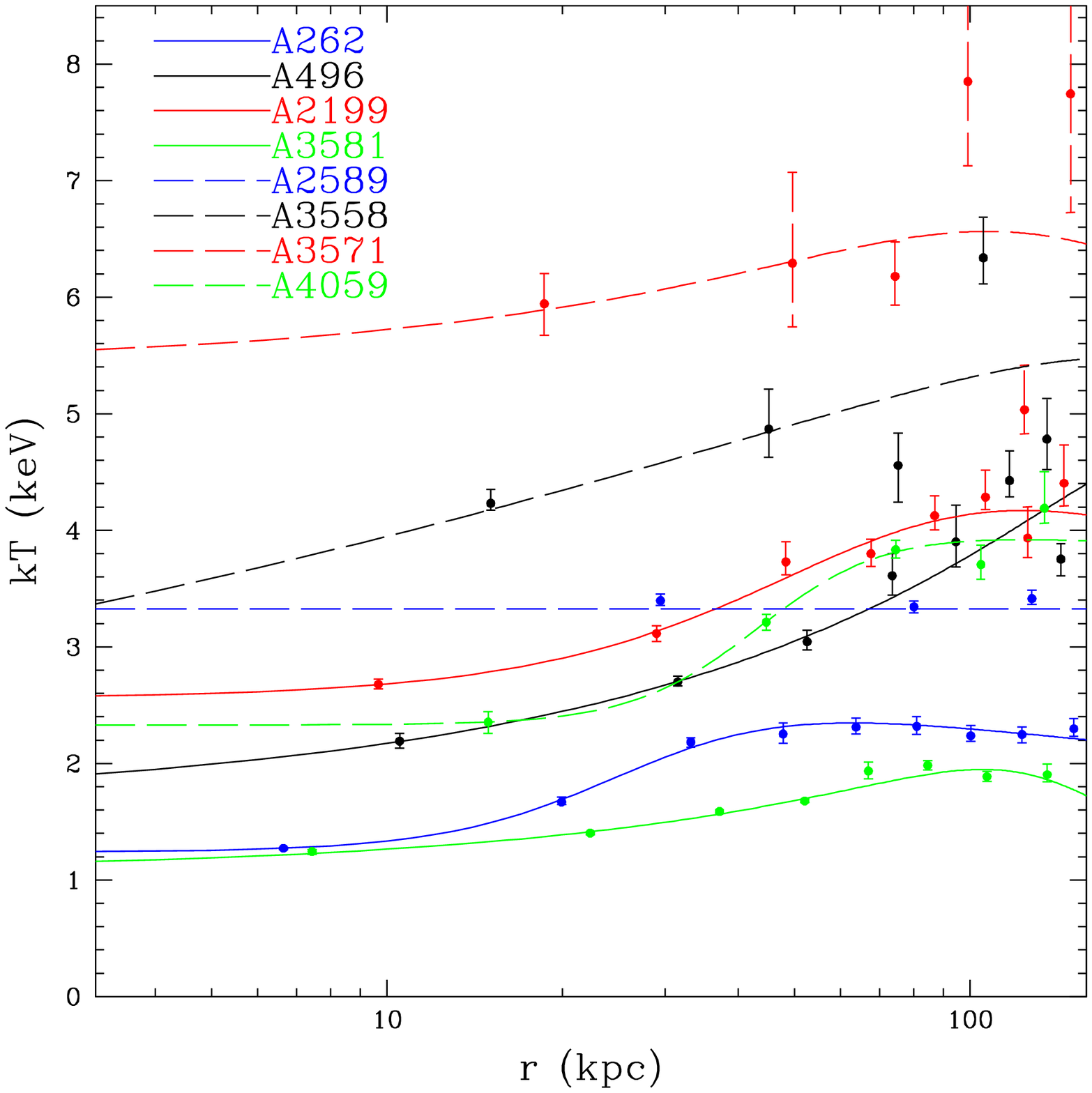}}
\caption{Deprojected temperature vs. radius for each cluster in our sample.
The line types are described in the caption to Fig. 1.}
\end{inlinefigure}

\bigskip

\noindent
time.  Clusters with short cooling times have the highest central densities
and lowest central entropies.  While all the clusters, 
except A2589, have 
a positive temperature gradient at small radii, the relative temperature 
decrement is greater in clusters with shorter cooling times.
To determine the total gas mass, Fe mass and Si mass within the
central 100~kpc in these clusters, we generated 1000 realizations of the cumulative
masses based on the best fit values of the gas density, Fe abundance and Si abundance
at each radius, along with their associated $1 \sigma$ uncertainties.  From these
1000 realizations we determined the mean cumulative mass at each radius along
with the $1 \sigma$ uncertainty.
The excess Fe mass within the central 100~kpc is calculated
assuming that all clusters have a uniform Fe abundance of 0.3 solar at large radii,
which is consistent with observations from ASCA, BeppoSax, XMM-Newton
and Chandra

\bigskip

\begin{inlinefigure}
\center{\includegraphics*[width=0.90\linewidth,bb=10 142 570 700,clip]{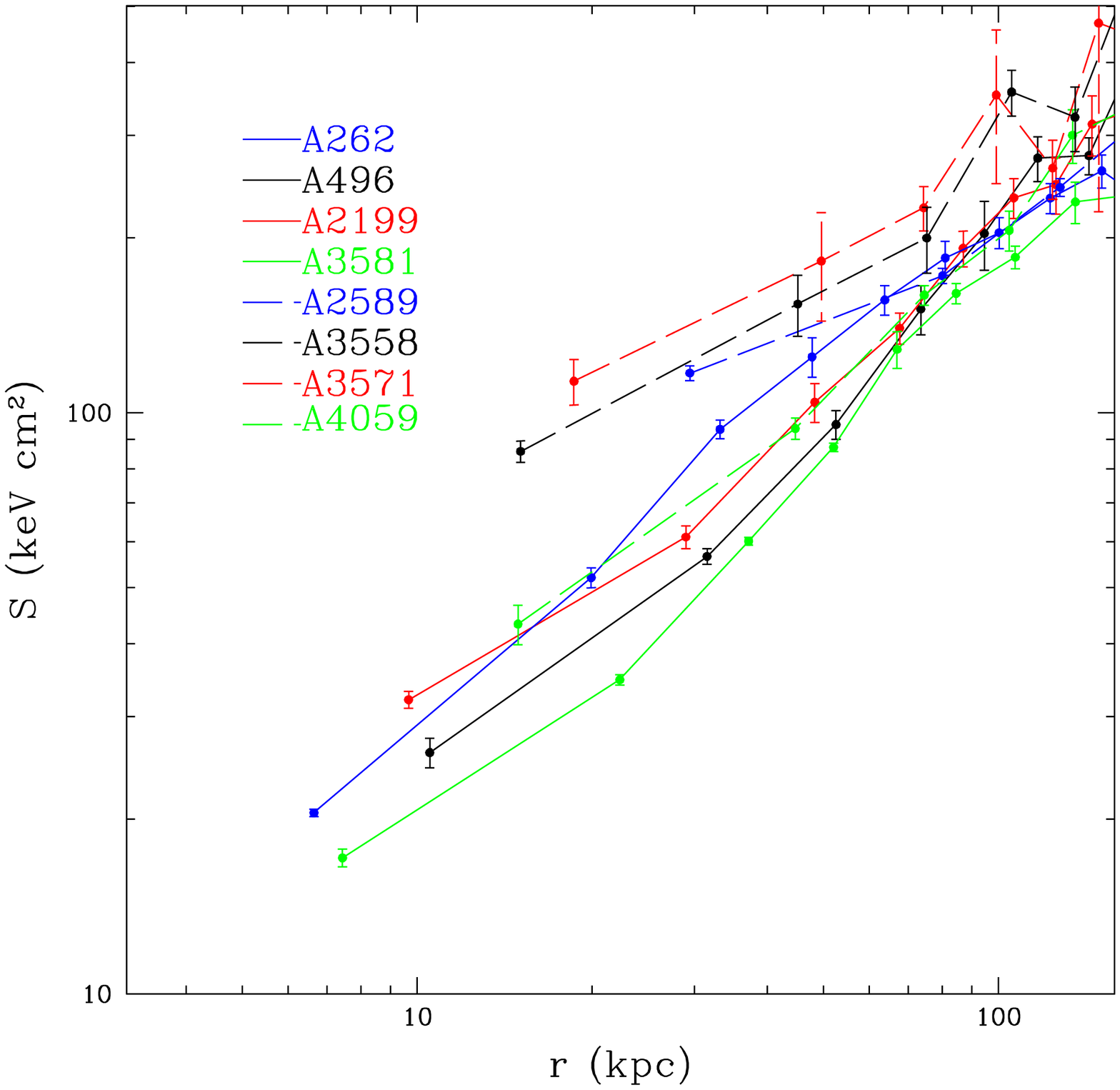}}
\caption{Entropy vs. radius for each cluster in our sample.
The line types are described in the caption to Fig. 1.}
\end{inlinefigure}

\begin{table*}[t]
\begin{center}
\caption{Gas Mass and Fe Mass Within the Central 100~kpc}
\begin{tabular}{lcccccc}
\hline
Name & $M_{gas}$ & $M_{Fe}(tot)$  & $M_{Fe}(ex)$ & $f_{Fe}(ex)$ & $M_{Fe}(tot)/L_R$ & $M_{Fe}(ex)/L_R$ \\
& ($M_{\odot}$) & ($M_{\odot}$) & ($M_{\odot}$) & & & \\
     \hline\hline
A262   & $2.1 \times 10^{11}$ & $(2.1 \pm 0.2) \times 10^{8}$ & $(1.3 \pm 0.1) \times 10^{8}$ & 0.62 & $1.3 \times 10^{-3}$ & $7.8 \times 10^{-4}$  \\
A496   & $5.1 \times 10^{11}$ & $(5.6 \pm 0.8) \times 10^{8}$ & $(3.6 \pm 0.5) \times 10^{8}$ & 0.64 & $1.3 \times 10^{-3}$ & $8.1 \times 10^{-4}$  \\
A2199  & $5.4 \times 10^{11}$ & $(4.8 \pm 0.5) \times 10^{8}$ & $(2.7 \pm 0.3) \times 10^{8}$ & 0.56 & $1.2 \times 10^{-3}$ & $6.6 \times 10^{-4}$  \\
A2589  & $3.8 \times 10^{11}$ & $(4.2 \pm 0.3) \times 10^{8}$ & $(2.8 \pm 0.2) \times 10^{8}$ & 0.67 & $1.5 \times 10^{-3}$ & $9.9 \times 10^{-4}$  \\
A3558  & $4.6 \times 10^{11}$ & $(4.3 \pm 0.9) \times 10^{8}$ & $(2.5 \pm 0.5) \times 10^{8}$ & 0.58 & $5.7 \times 10^{-4}$ & $3.3 \times 10^{-4}$  \\
A3571  & $5.8 \times 10^{11}$ & $(4.8 \pm 0.1) \times 10^{8}$ & $(2.6 \pm 0.5) \times 10^{8}$ & 0.54 & $5.5 \times 10^{-4}$ & $3.0 \times 10^{-4}$  \\
A3581  & $2.4 \times 10^{11}$ & $(2.2 \pm 0.1) \times 10^{8}$ & $(1.3 \pm 0.08) \times 10^{8}$& 0.59 & $1.3 \times 10^{-3}$ & $7.6 \times 10^{-4}$  \\
A4059  & $5.0 \times 10^{11}$ & $(6.4 \pm 0.6) \times 10^{8}$ & $(4.5 \pm 0.5) \times 10^{8}$ & 0.70 & $1.2 \times 10^{-3}$ & $8.8 \times 10^{-4}$  \\
\hline
\end{tabular}

\end{center}
\noindent
Notes:  Cluster name, gas mass within 100~kpc, total Fe mass within 100~kpc,
excess Fe mass within 100~kpc, ratio of the excess Fe mass to the total Fe mass,
ratio of total Fe mass to $L_R$ of the CDG
and ratio of excess Fe mass to $L_R$ of the CDG.  All errors are given at the $1 \sigma$
confidence level.
\end{table*}

\bigskip

\noindent 
(e.g., Finoguenov, David \& Ponman 2000;
De Grandi et al. 2004; Tamura et al. 2004; Vikhlinin et al. 2005;
Baldi et al. 2007).  The resulting gas mass, total Fe mass, excess Fe mass, ratio
of excess Fe mass to total Fe mass and the ratios of total and excess Fe mass to
the R-band luminosity of the CDGs are shown in Table 2.
The mean values of $M_{Fe}/L_R$ and $M_{Fe}(ex)/L_R$ within the central 100~kpc
in our sample are $9.0 \times 10^{-4} M_{\odot}/L_R$ 
and $6.9 \times 10^{-4} M_{\odot}/L_R$, respectively.
Using an average B-R=1.8 for the CDGs in our sample (from the photometric
tables in NED), gives mean values of
$M_{Fe}/L_B$ and  $M_{Fe}(ex)/L_B$ = 0.0015 and $0.0012 M_{\odot}/L_B$.
Our mean value of $M_{Fe}/L_B$ within the central 100~kpc is consistent
with the results in Finoguenov et al. (2000) derived from a sample of 11
relaxed clusters observed by ASCA.

The total Si mass within the central 100~kpc in these clusters is shown in
Table 3 along with the total Si-to-Fe mass ratio relative to the solar value.
In addition to the statistical errors shown for the Si mass, a systematic
error of 10\% should also be

\bigskip

\begin{inlinefigure}
\center{\includegraphics*[width=0.90\linewidth,bb=10 142 570 700,clip]{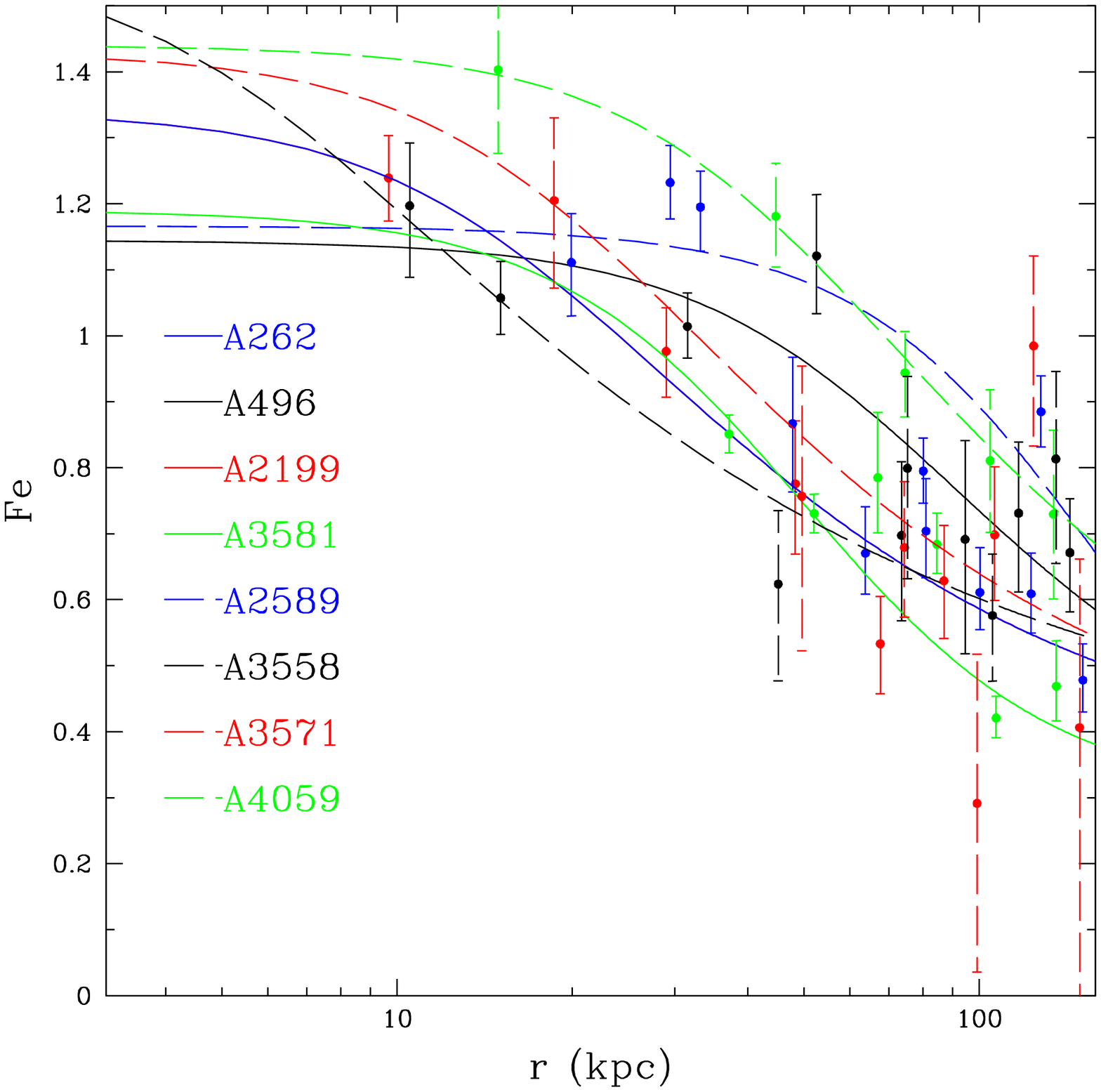}}
\caption{Fe abundance (relative to the solar value in Anders \& Grevesse 1989)
vs. radius for each cluster in our sample.
The line types are described in the caption to Fig. 1.}
\end{inlinefigure}

\noindent
included to account for cross-calibration
issues between the EPIC cameras (de Plaa et al. 2007).
All of the [Si/Fe] values 
in our cluster sample are subsolar, indicating a
greater relative enrichment by SNe Ia.
The mean value of [Si/Fe] for our cluster 
sample is -0.17.
Adopting the yields for SNe Ia from from the delayed detonation model (WDD2)
of Iwamoto et al. (1999)
($y_{SNe Ia}(Fe)=0.713~M_{\odot}$ and $y_{SNe Ia}(Si)=0.206~M_{\odot}$)
and the yields for core collapse supernovae (SNcc) from Tsujimoto et al. (1995)
($y_{cc}(Fe)=0.084~M_{\odot}$ and $y_{cc}(Si)=0.105~M_{\odot}$),
the mean value of [Si/Fe] in our sample gives a SNe Ia fraction
of $f_{SNe Ia} = 0.53$.  Tsujimoto et al. (1995) estimated that
$N_{SNe Ia}/N_{cc}=0.15$ ($f_{SNe Ia} = 0.17$) is required to reproduce
the observed abundance pattern of 14 elements from oxygen to nickel in the
Galaxy based on the SNe Ia yields from the W7 model of Thielemann,
Nomoto \& Hashimoto (1993), the same SNcc
yields that we have adopted and the solar abundances in
Anders \& Grevesse (1989).  Based on our adopted
supernova yields and the solar abundances in Grevesse \& Sauval (1998),
$f_{SNe Ia} = 0.23$ is required to reproduce the solar Si-to-Fe mass ratio.

Our estimates of $f_{SNe Ia}$ are roughly consistent with previous studies based on
XMM-Newton observations of clusters.  de Plaa et al. (2006) obtained
$f_{SNe Ia} = 0.38-0.50$ for Sersic 159-03.
Werner et al. (2006) obtained a lower value of $f_{SNe Ia} = 0.27 \pm 0.03$
based on a similar analysis of a deep XMM-Newton observation of 2A0335+09.
In a recent paper, de Plaa et al. (2007) demonstrated that
the derived value of $f_{SNe Ia}$ can vary from 0.22 to 0.72 depending on the
adopted supernova yields, progenitor metallicity and stellar initial mass function.
Our results, together with previous studies, all indicate that the
gas surrounding the CDG in clusters has experienced a greater
relative enrichment from SNe Ia compared to solar abundance gas.

Since the Fe yield from SNe Ia is almost 10 times greater than
the Fe yield from SNcc, the time required to accumulate
the observed excess Fe mass can be estimated assuming that only SNe Ia from the
CDG are responsible for the enrichment.
Scaling the present day SNe Ia rate in early-type galaxies given
in Cappellaro et al. (1999), using an average B-R=1.8 for the CDGs
in our clusters and our adopted cosmology gives,

\begin{equation}
R_{SNe Ia} = 1.63 \times 10^{-2} \left( {L_R} \over {10^{11} L_{\odot R}} \right)~yr^{-1}.
\end{equation}

\begin{table*}[t]
\begin{center}
\caption{Si Mass Within the Central 100~kpc and Inferred SNe Ia Heating Rate}
\begin{tabular}{lcccccc}
\hline
Name & $M_{Si}(tot)$ & $[Si/Fe]_{tot}$ & $\tau_0$ & $\tau_e$ & $H_{SNeIa}$ & $H_{SNeIa}/L_{bol}$ \\
& ($M_{\odot}$) & & (Gyr) & (Gyr) & (erg~s$^{-1}$) & \\
     \hline\hline
A262   & $(9.4 \pm 1.5) \times 10^7$ & $-0.10 \pm 0.07$ & 6.7 & 5.4 & $8.7 \times 10^{41}$ & 0.04 \\
A496   & $(2.7 \pm 0.8) \times 10^8$ & $-0.07 \pm 0.11$ & 7.0 & 5.5 & $2.3 \times 10^{42}$ & 0.02 \\
A2199  & $(2.0 \pm 0.6) \times 10^8$ & $-0.13 \pm 0.12$ & 5.6 & 4.7 & $2.1 \times 10^{42}$ & 0.01 \\
A2589  & $(1.1 \pm 0.3) \times 10^8$ & $-0.30 \pm 0.10$ & 8.6 & 6.4 & $1.5 \times 10^{42}$ & 0.03 \\
A3558  & $(2.0 \pm 0.6) \times 10^8$ & $-0.08 \pm 0.11$ & 2.9 & 2.6 & $3.9 \times 10^{42}$ & 0.04 \\
A3571  & $(1.8 \pm 0.5) \times 10^8$ & $-0.17 \pm 0.08$ & 2.6 & 2.4 & $4.4 \times 10^{42}$ & 0.02 \\
A3581  & $(7.7 \pm 0.7) \times 10^7$ & $-0.21 \pm 0.06$ & 6.5 & 5.2 & $8.9 \times 10^{41}$ & 0.03 \\
A4059  & $(1.7 \pm 0.4) \times 10^8$ & $-0.33 \pm 0.10$ & 7.6 & 5.8 & $2.7 \times 10^{42}$ & 0.03 \\
\hline
\end{tabular}

\end{center}
\noindent
Notes:  Cluster name, total Si mass within the central 100~kpc, ratio of total Si mass
to Fe mass relative to the solar ratio within 100~kpc, the time required
to produce the presently observed excess Fe mass assuming a constant
SNe Ia rate ($\tau_0$) and an evolving SNe Ia rate ($\tau_e$),
the SNe Ia heating rate and the SNe Ia heating rate relative to the
bolometric X-ray luminosity of the gas.  All errors are given at the $1 \sigma$
confidence level.
\end{table*}

\bigskip

\noindent
The SNe Ia rate for E/S0 galaxies derived by Mannucci et al. (2005) is
approximately 20\% lower than the rate in Cappellaro et al. (1999),
but is within the statistical uncertainties.
Table 3 shows the time, $\tau_0$, required to reproduce the observed excess Fe
mass in these clusters using the Fe yield for SNe Ia given above and
assuming that $R_{SNe Ia}$ is a constant.  Accounting for Fe enrichment
from SNcc using $f_{SNe Ia} = 0.53$, the mean value in our sample, only
decreases the required enrichment time by 11\%. We also estimate
the time required to reproduce the excess Fe mass if $R_{SNe Ia} \propto (t/t_H)^{-1}$
(where $t_H=13.7$~Gyr).  This time-dependence is
consistent with previous calculations for the evolving rate of
SNe Ia (e.g., Loewenstein \& Mathews 1987; David, Forman \& Jones 1990; Renzini et al. 1993).
Table 3 shows that the excess Fe mass within the central 100~kpc in these clusters
can be produced by SNe Ia from the CDGs in 3-7~Gyr.  These times are lower limits
since we do not account for the expulsion of Fe beyond the central 100~kpc or
mass deposition.

Assuming that each SNe Ia generates $10^{51}$~erg, the
present day SNe Ia heating rate is only a few percent of the bolometric
X-ray luminosity within the central 100~kpc in these clusters (see Table 3).
Core collapse supernovae are unlikely to significantly increase the
total supernova heating rate since the observed star formation rate in
the CDGs in cool-core cluster seldom exceeds $20~M_{\odot}$~yr$^{-1}$
(Rafferty et al. 2006).  In A262, the star formation rate is constrained to be less than
$0.015~M_{\odot}$~yr$^{-1}$.  Assuming a Salpeter IMF (for which 1 SNcc
is produced for every $100 M_{\odot}$ of gas consumed into stars)
and a star formation rate of $10~M_{\odot}$~yr$^{-1}$,
the heating rate from SNcc would be $3.0 \times 10^{42}$~erg~s$^{-1}$, which would
significant increase the heating rate from SNe Ia alone.

\subsection{The Relative Distribution of the Excess Fe}

If the Fe produced by SNe Ia remains stationary after being
injected into the hot gas and there is no mass deposition, then the ratio
of excess Fe mass density to stellar mass density
($\rho_{Fe}(ex)/\rho_*$) should be independent of radius.
This prediction is in strong conflict with the observed trend which clearly
shows that the excess Fe is significantly more extended than the
stellar mass in each cluster (see Fig. 5). The curves in Fig. 5
are calculated using a cubic spline fit to the Fe abundance profiles.
Fig. 5 also shows that there is some evidence that the excess Fe is more 
centrally concentrated in clusters with shorter cooling times.
For comparison, we also show in Fig. 5 the expected values of
$\rho_{Fe}(ex)/\rho_*$ if the Fe has accumulated over the past
5 or 7.5~Gyr (using the evolving SNe Ia rate given above)
and remained stationary without any mass deposition.
While the average values of $\rho_{Fe}(ex)/\rho_*$ within the central
100~kpc are consistent with the expected accumulation of Fe over
the past 5-7.5~Gyr, there is clearly a deficit of Fe within the
central 30-50~kpc and an excess at larger radii.

\bigskip

\begin{inlinefigure}
\center{\includegraphics*[width=0.90\linewidth,bb=10 142 570 700,clip]{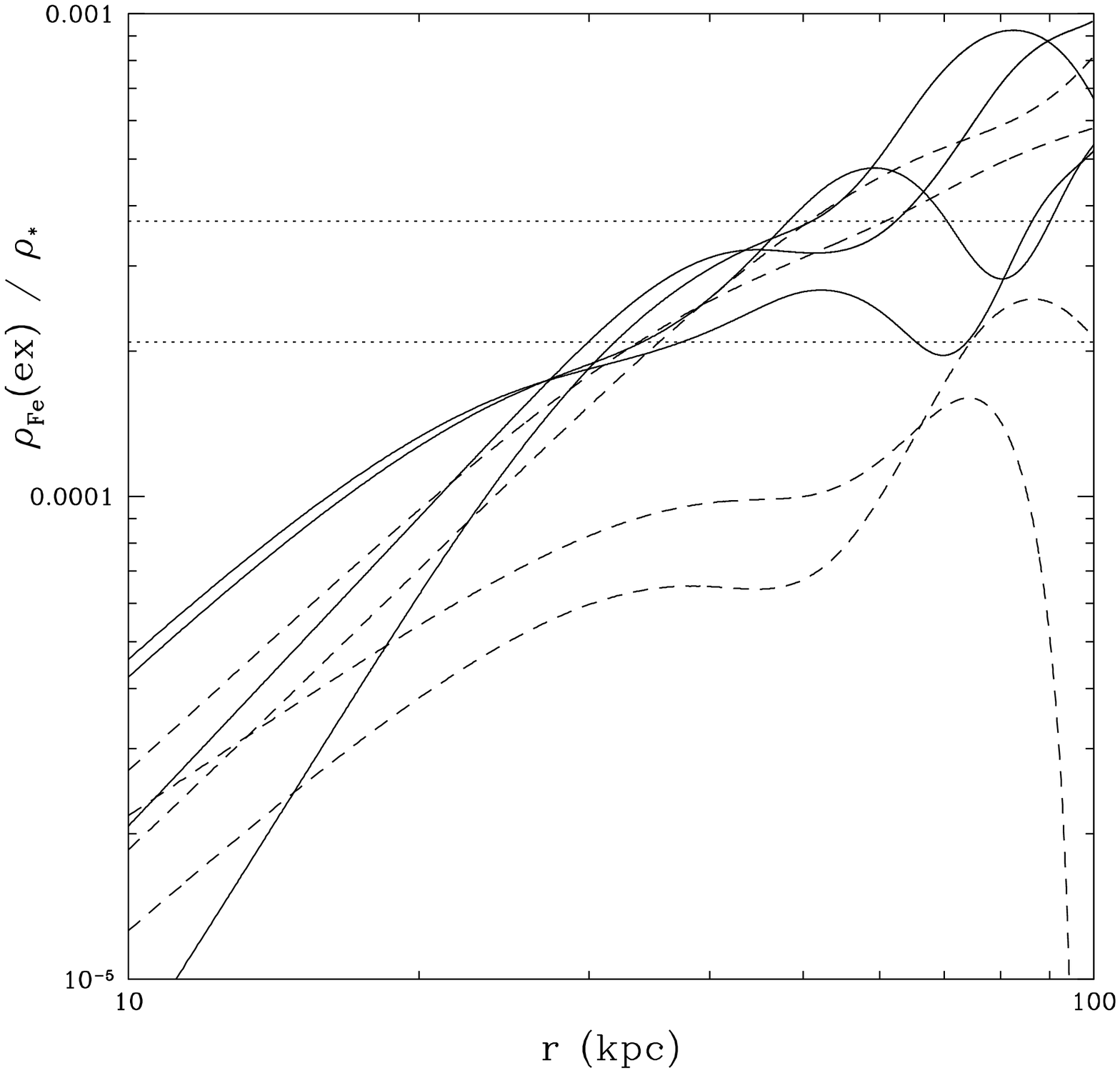}}
\caption{Ratio of excess Fe mass density to stellar mass density vs. radius for
each cluster in our sample.  The line types are described in the caption to Fig. 1.
The two horizontal dotted lines indicate the ratio
that would result from the accumulation of Fe from SNe Ia
over the past 5 (lower line) and 7.5 Gyr (upper line) assuming
the Fe remains stationary after being injected and no mass deposition.
From top to bottom,
the clusters with short cooling times are: A2199, A496, A262 and A3581.
From top to bottom, the clusters with long cooling times are: A4059, A3571, A2589
and A3558.  The deprojected Fe abundance in A3571
drops slightly below 0.3 solar near 100~kpc which produces a sharp
decline in $\rho_{Fe}(ex)/\rho_*$ at this radius.}
\end{inlinefigure}

\begin{inlinefigure}
\center{\includegraphics*[width=0.90\linewidth,bb=10 142 570 700,clip]{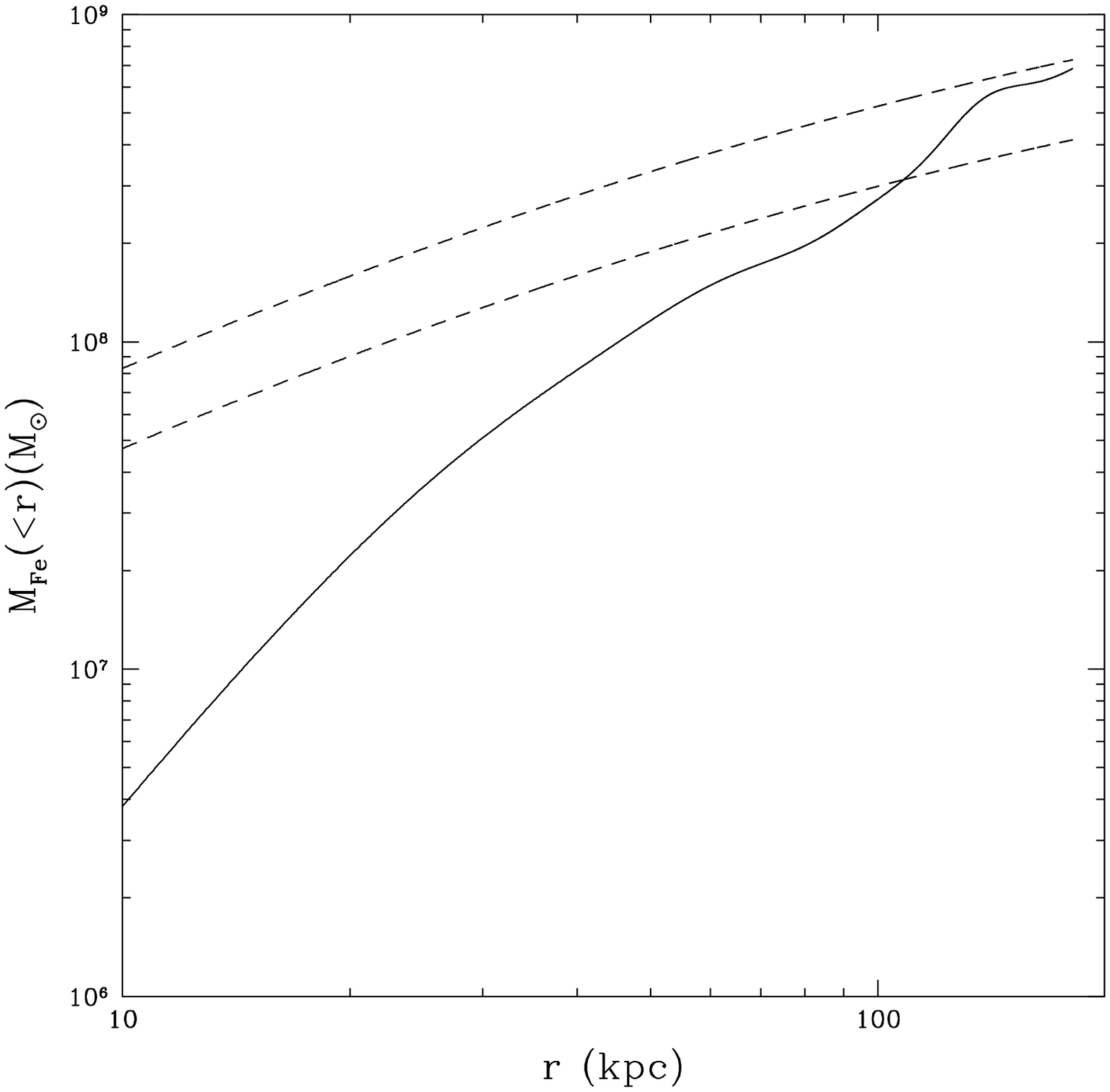}}
\caption{Integrated excess Fe mass in A2199 (solid line) compared to predictions based on
the accumulation of Fe from SNe Ia over the past 5 (lower dashed line)
and 7.5~Gyr (upper dashed line) assuming the Fe remains stationary after being
injected and no mass deposition.}
\end{inlinefigure}

\section{Mechanisms for Producing an Extended Fe Distribution}

We explore two possible explanations for the more extended
distribution of the excess Fe relative to the stellar mass, 
including: 1) bulk expansion of the gas 
due to AGN heating and 2) turbulent diffusion of Fe.

\subsection{Bulk Expansion}

The integrated excess Fe mass observed in A2199 and the predicted values if
the Fe has accumulated from SNe Ia over the past 5 or 7.5~Gyr and remained stationary
without mass deposition are shown in Fig. 6.  This figure shows that the Fe content
of the hot gas within the central 30~kpc is a factor of 10 less than the predicted value.
We can estimate how much the excess Fe must have expanded since being injected 
into the ICM by SNe Ia by comparing the predicted and observed
integrated excess Fe masses shown in Fig. 6.
For example, assuming that the excess Fe has accumulated from SNe Ia over the 
past 5~Gyr in A2199 and remained stationary without any mass deposition, 
there should be $10^8~M_{\odot}$ of excess Fe within the central 17~kpc.  The observed 
Fe distribution shows that there is $10^8~M_{\odot}$ of excess Fe within 
the central 45~kpc, thus the Fe must have undergone a net expansion 
from 17~kpc to about 45~kpc during this time.   This estimate does not include
the Fe injected at radii between 17 and 45~kpc while the gas is expanding,
and is thus a lower limit on the net expansion.

Using this method, we can compute the minimum net expansion factor
as a continuous function of radius for each cluster using
the cubic spline fits to the Fe abundance profiles.
Fig. 7 shows the required expansion factor $(r_f-r_i) / r_i$, where $r_i$ is the
initial radius and $r_f$ is the final (present) radius, as a function 
of $r_i$ for each cluster.
The curve with the largest wiggles in Fig. 7 corresponds to 
A3571, which has the poorest photon statistics 
in our sample, and hence, the greatest uncertainties in the Fe abundance.
Fig. 7 shows some evidence that the net expansion factor of the excess Fe may 
be greater in clusters with longer central cooling times.

\begin{inlinefigure}
\center{\includegraphics*[width=0.90\linewidth,bb=10 142 570 700,clip]{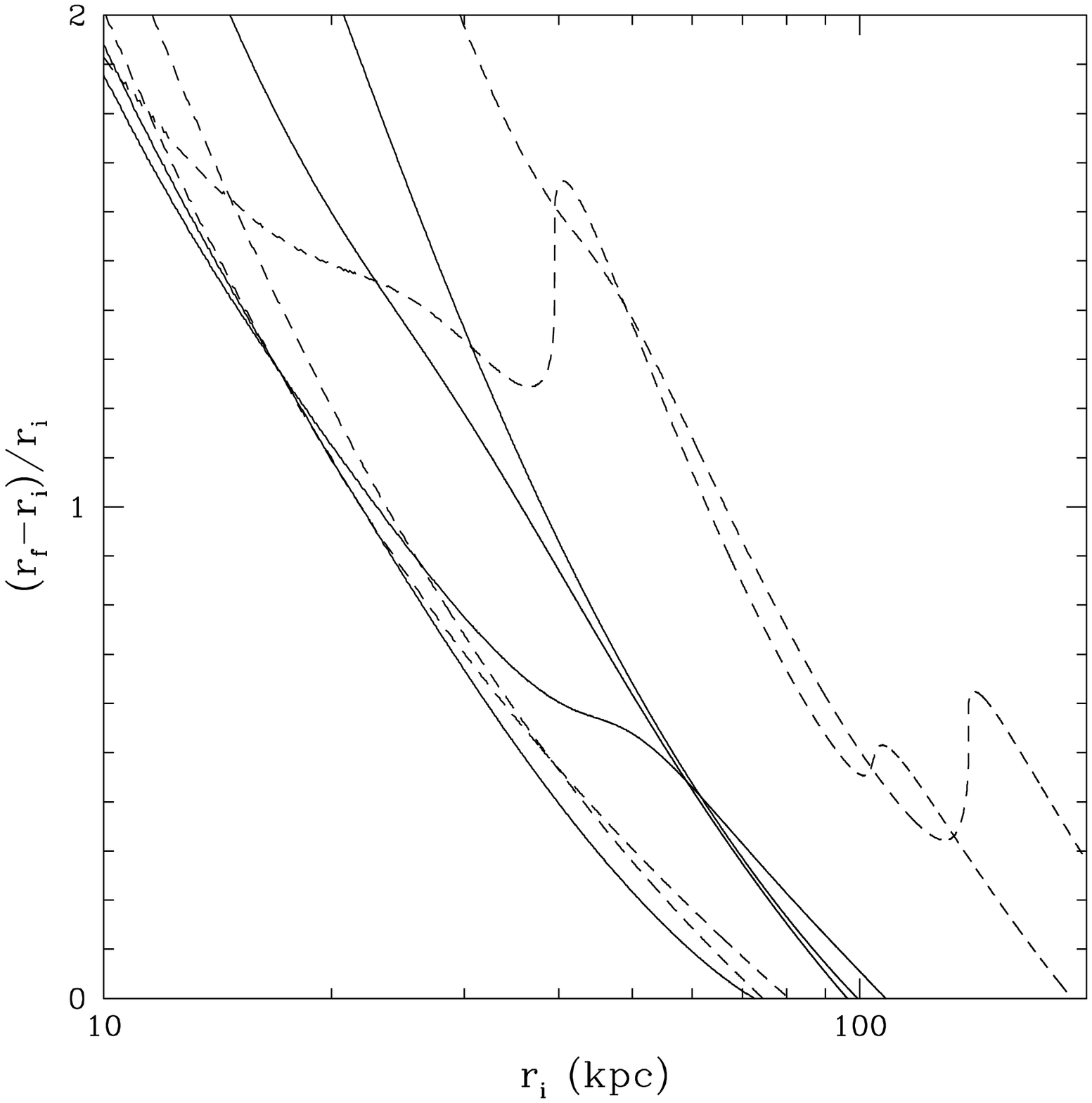}}
\caption{The required expansion factor of the gas to reproduce the observed integrated excess Fe
mass for all of the clusters in our sample assuming the Fe has accumulated from SNe Ia over
the past 5~Gyr and no mass deposition. The line types are described in the caption to Fig. 1.}
\end{inlinefigure}

\bigskip

If we assume that the excess Fe and gas expand together, then the energy required 
to inflate the gas is:

\begin{equation}
\Delta E = \Delta W + \Delta U 
\end{equation}

\noindent
where $\Delta W$ is the difference in the gravitational potential energy
between the final and initial states of the gas and
$\Delta U$ is the difference in thermal energy.  
We compute the gravitating mass distribution in each cluster 
using the deprojected gas density profile shown in Fig. 1, a parametric 
fit to the deprojected temperature profile based on the functional
form used by Allen, Schmidt \& Fabian (2001), and the
assumption of hydrostatic equilibrium. The potential energy difference 
for a given mass shell
is $\Delta W = \Delta M_{gas}(r_f)(\phi(r_f)-\phi(r_i))$, where
$\Delta M_{gas}(r_f)$ is the gas mass within a spherical shell centered at radius
$r_f$ and $\phi$ is the gravitational potential.  Assuming that the gas is in 
hydrostatic equilibrium both initially and at present,
we show in the Appendix that the change in thermal energy within 
a given mass shell can be written as:

\begin{equation}
\Delta U =  {{1} \over {2}} (v_k^2(r_f) - v_k^2(r_i))\Delta M_{gas}(r_f)
\end{equation}

\noindent
where $v_k$ is the Keplerian velocity.  

The cumulative total energy difference
between the final and initial states of the gas are shown in Fig. 8 for each
cluster based on the expansion factors in Fig. 7.  We do not include the surface
term (i.e., the PdV work done by the expanding gas) in the total energy since,
as shown in the Appendix, this term approaches zero at large radii.
Fig. 8 shows that $10^{60} - 10^{61}$~erg must
have been deposited within the central 100~kpc of these clusters within the past 5~Gyr
to inflate the Fe to its present distribution assuming that the Fe and
gas expand together.   The total potential and thermal energy differences
for each cluster within the central 100~kpc are shown in Table 4.  This table
shows that the potential energy difference accounts for 70-80\% of the
total energy difference.  Our computed energy differences are comparable to the 
AGN mechanical energy required to evacuate the X-ray cavities and drive the 
shocks observed in 

\begin{table*}[t]
\begin{center}
\caption{Energy Requirements for Bulk Expansion of the Gas}
\begin{tabular}{lccc}
\hline
Name & $\Delta W$ & $\Delta U$ & $\Delta E$ \\
& (erg) & (erg) & (erg) \\
     \hline\hline
A262   & $9.1 \times 10^{59}$ & $2.1 \times 10^{59}$ & $1.1 \times 10^{60}$ \\
A496   & $9.9 \times 10^{59}$ & $2.2 \times 10^{59}$ & $1.2 \times 10^{60}$ \\
A2199  & $2.9 \times 10^{60}$ & $8.0 \times 10^{59}$ & $3.7 \times 10^{60}$ \\
A2589  & $5.8 \times 10^{59}$ & $3.0 \times 10^{59}$ & $8.8 \times 10^{59}$ \\
A3558  & $1.1 \times 10^{61}$ & $2.1 \times 10^{60}$ & $1.3 \times 10^{61}$ \\
A3571  & $1.6 \times 10^{61}$ & $1.2 \times 10^{60}$ & $1.7 \times 10^{61}$ \\
A3581  & $1.0 \times 10^{60}$ & $2.0 \times 10^{59}$ & $1.2 \times 10^{60}$ \\
A4059  & $7.7 \times 10^{59}$ & $5.9 \times 10^{59}$ & $1.4 \times 10^{60}$ \\
\hline
\end{tabular}

\end{center}
\noindent
Notes:  Cluster name, increase in potential energy, thermal energy
and total energy required to reproduce the observed Fe profiles
assuming bulk inflation of the gas within the central 100~kpc.
\end{table*}

\bigskip

\noindent
cool-core clusters (e.g., McNamara et al. 2000, 
Fabian et al. 2003, Blanton et al. 2003, Nulsen et al. 2005a, Nulsen et al. 2005b;
Forman et al. 2005, Wise et al. 2006).  The average power required
to inflate the gas within the central 100~kpc is $10^{43} - 10^{44}$~erg~s$^{-1}$, which 
is also comparable to the mean ``cavity power" derived from a sample of cool-core
clusters analyzed by Birzan et al. (2004) and Rafferty et al. (2006).

The energy required to reproduce the observed Fe distributions shown in Fig. 8 
is a lower limit on the total mechanical energy generated by an AGN,
since some of the energy will be radiated 
away and some of the energy will be deposited beyond 100~kpc. For example, the 
shock in Hydra A is located 200-300~kpc from the center of the cluster 
(Nulsen et al. 2005a).  
Fig. 9 shows the time it will take to dissipate the additional energy required 
to inflate the gas within a given radius, $t_{loss}=\Delta E(<r)/L_{bol}(<r)$.  
This figure shows that the excess energy within the 
central 30~kpc of these clusters will be dissipated within
$1-3 \times 10^8$~yr.  If AGN outbursts repeat on approximately this timescale, 
then only the gas within the central 30~kpc will be able to cool and contract before
the next AGN outburst, while gas beyond this radius will 
continuously expand.  The gas mass within the central 30~kpc is typically 10\% 
of the total gas mass within the central 100~kpc in these clusters.
Thus, if AGN outbursts repeat on timescales of approximately $1-3 \times 10^8$~yr,
then only 10\% of the total gas mass within the central 100~kpc is available
for further cooling and accretion onto the central supermassive black hole.

\subsection{Turbulent Diffusion of Iron}

AGN inflated X-ray cavities are commonly found in the centers of 
cool-core clusters. As these cavities buoyantly rise outward,
they will generate turbulent gas motions in their wake. AGN driven
shocks expanding into inhomogeneous cluster gas can also 
generate turbulent gas motions (Heinz et al. 2006) as well as
cluster mergers and the motion of the host galaxies in a cluster. Based on the high
optical depth toward the center of the Perseus cluster and the lack
of any observed resonant scattering in the He-like Fe K$_{\alpha}$ 
to K$_{\beta}$ line ratio, Churazov et al. (2004) concluded that 
there must be significant gas turbulence in the central region of the cluster.
Such turbulence will also produce an outward diffusion of Fe as higher abundance
gas at small radii is mixed with lower abundance gas at large radii.

We model the outward diffusion of Fe using the same technique used
by Rebusco et al. (2005) and (2006) who estimated the diffusion
coefficient, $D$, and dissipative heating rate in Perseus and 7 other 
groups and clusters.  They performed a series of time-dependent simulations 
for each cluster with a range of values of $D$. Each simulation started 
with a uniform Fe abundance. 
The excess Fe was assumed to originate from SNe Ia with the same
distribution as the stars in the CDG and then diffuse outward.  They found that
a diffusion coefficient of $D \approx 2 \times 10^{29}$~cm$^2$~s$^{-1}$ is
required to reproduce the presently observed Fe abundance profiles in these 
groups and clusters.  
This method was also used by Graham et al. (2006) to estimate the diffusion
coefficient in the Centaurus cluster and determine the characteristics of the 
turbulence required to locally balance radiative cooling.
In addition, Roediger et al. (2007) performed 
a series of numerical simulations to study
the effects of buoyantly rising bubbles on the abundance gradients in cluster
atmospheres.  They found that the dredging up of low entropy,

\bigskip

\begin{inlinefigure}
\center{\includegraphics*[width=0.90\linewidth,bb=10 142 570 700,clip]{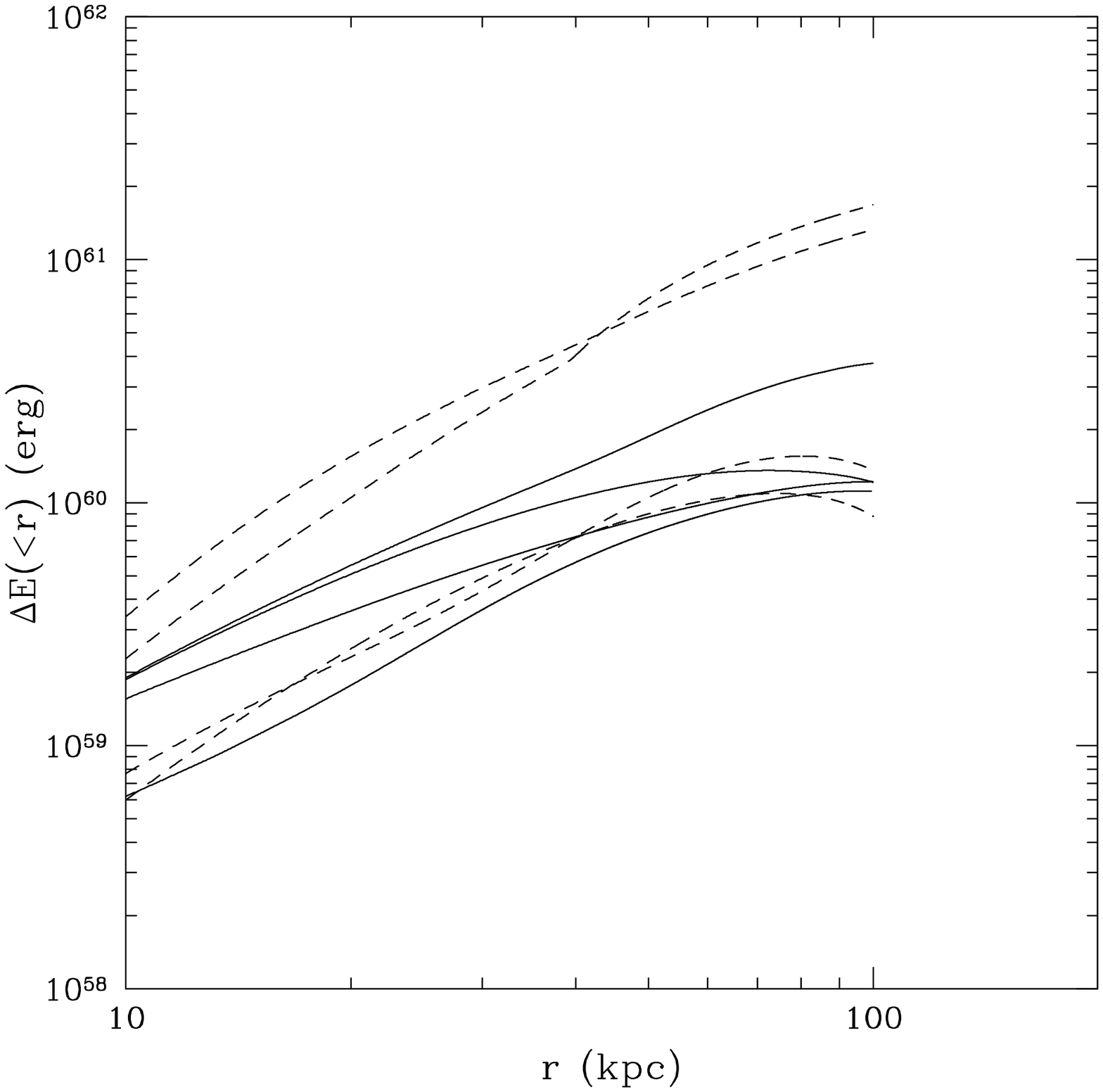}}
\caption{The integrated total energy required to inflate the
gas and reproduce the observed integrated excess Fe mass distribution in each cluster in our sample
assuming the Fe has accumulated from SNe Ia over the past 5~Gyr.
The line types are described in the caption to Fig. 1.}
\end{inlinefigure}

\begin{inlinefigure}
\center{\includegraphics*[width=0.90\linewidth,bb=10 142 570 700,clip]{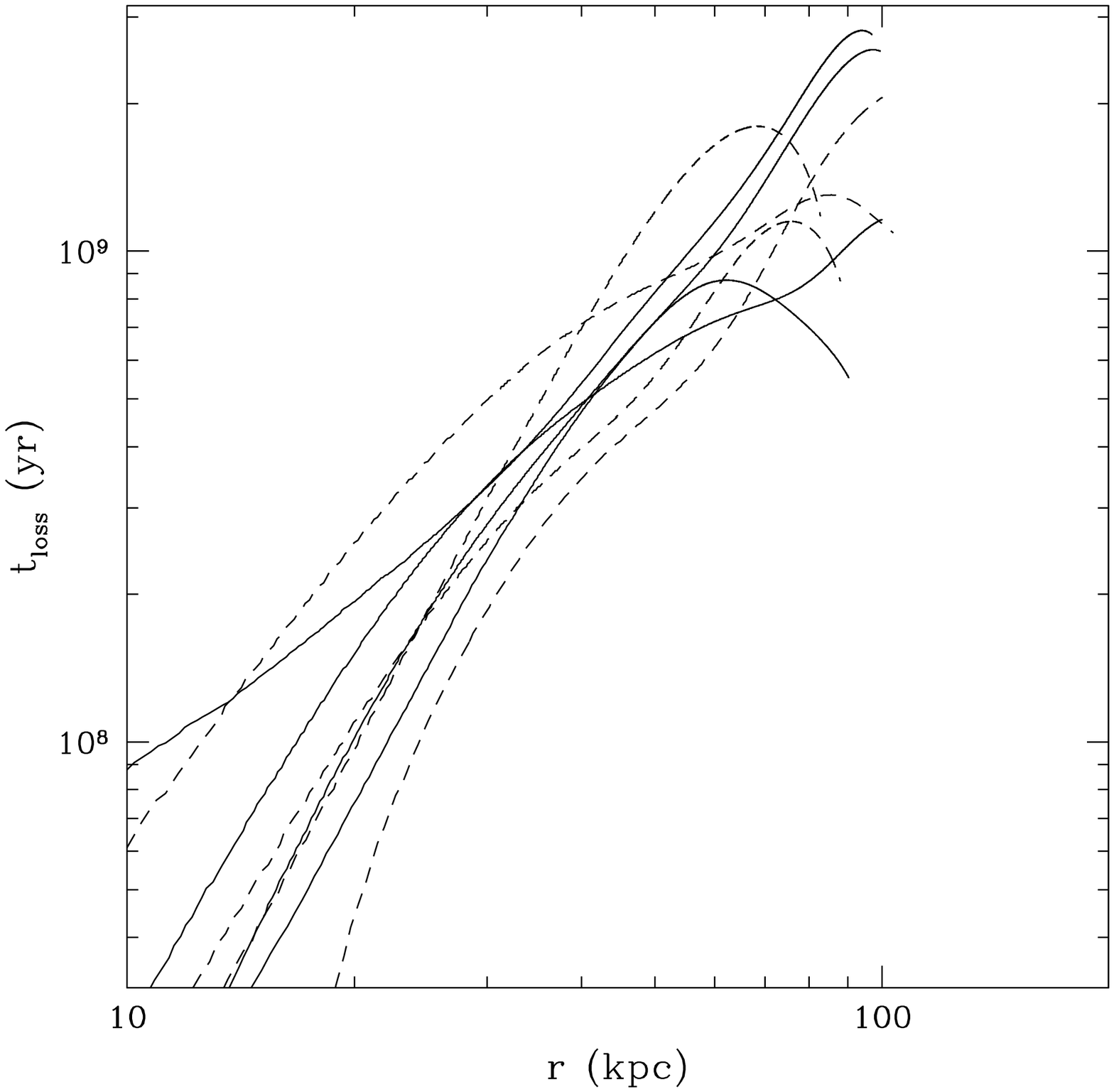}}
\caption{The time to dissipate the excess energy shown in Fig. 8 through radiative losses.
The line types are described in the caption to Fig. 1.}
\end{inlinefigure}

\bigskip

\noindent
high abundance gas in the 
wake of buoyant bubbles can produce effective diffusion coefficients comparable to that 
required to reproduce the observed Fe abundance profiles in clusters.

Assuming the gas density profile is time-independent, there is no bulk motion of the gas,
and isotropic turbulence, the Fe diffusion equation can be written as:

\begin{equation}
{{\partial (\rho_g Z_{Fe})} \over {\partial t}} = \vec{\nabla} \cdot (D \rho_g \vec{\nabla}Z_{Fe} ) + \dot{\rho}_{Fe,inj}
\end{equation}

\noindent
where $\rho_g$ is the gas mass density, $Z_{Fe}$ is the Fe abundance by mass
and $\dot{\rho}_{Fe,inj}$ is the Fe mass injection rate per unit volume
from SNe Ia.  
Assuming steady-state (i.e., the outward mass flux of Fe across a given surface 
is equal to the mass injection rate of Fe within that surface), the diffusion
equation can be written as:

\begin{equation}
\int D \rho_g \vec{\nabla}Z_{Fe} \cdot \vec{dA} =  - \dot{M}_{Fe}(<R)
\end{equation}

\noindent
Fig. 10 shows the required diffusion coefficient to maintain a 
steady-state Fe abundance profile as a function of radius for 
each cluster in our sample.  For most clusters,  the diffusion
coefficient decreases outward, attaining a minimum value between 20-50 kpc,
and then increases at larger radii.  This results from the flattening of 
the Fe abundance profiles at small and large radii.
As the Fe abundance gradient decreases, larger diffusion coefficients are required to 
transport the Fe outward at the same rate that it is being 
injected by SNe Ia.  

The smallest values of $D$ occur in regions
where the Fe abundance profiles are the steepest.

We can determine if the outward diffusion of Fe is primarily driven by 
particle diffusion or turbulent gas motions by comparing
the diffusion coefficient derived above with the particle diffusion 
coefficient of Fe, given by: $\kappa_{Fe} \approx 1/3 \lambda_{Fe}v_{Fe}$,
where $\lambda_{Fe}$ is the Fe mean free path 
and $v_{Fe}$ is the thermal Fe velocity.  The shortest Fe mean free path 
is the Fe-proton mean free path. Using the expressions in Spitzer (1962) 
to compute the Fe-proton mean free path assuming that all Fe is He-like (i.e., an 
ion charge of 24), we plot the ratio $D/\kappa_{Fe}$ as a function of 

\bigskip

\begin{inlinefigure}
\center{\includegraphics*[width=0.90\linewidth,bb=10 142 570 700,clip]{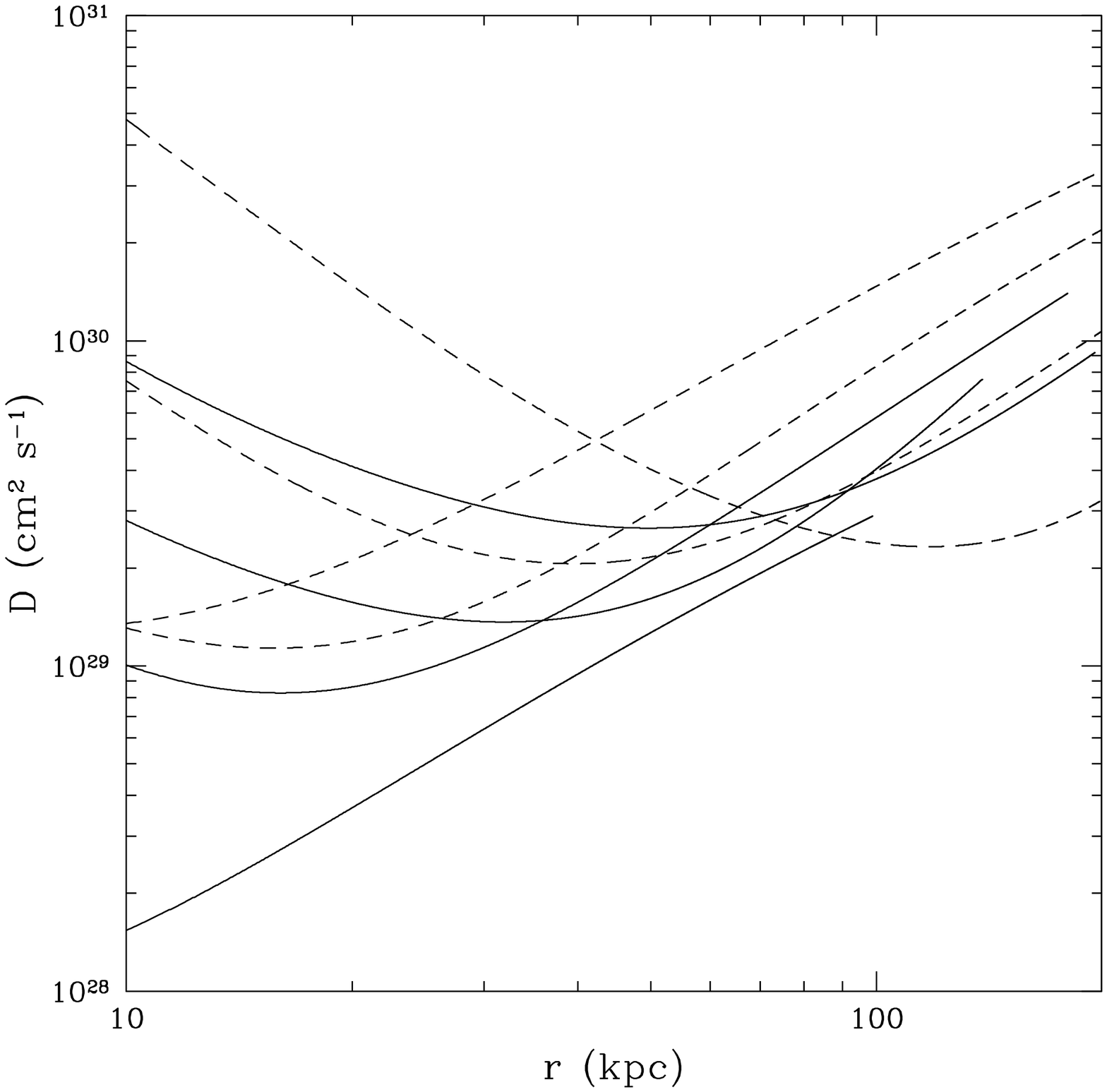}}
\caption{The diffusion coefficient, $D$, required to maintain a steady-state Fe abundance
profile in each cluster in our sample.  The line types are described in
the caption to Fig. 1.}
\end{inlinefigure}

\noindent
radius for each cluster in Fig. 11. This figure shows that the 
required rate of Fe diffusion to maintain a steady-state Fe abundance 
profile is $10^3 - 10^4$ times greater than the particle diffusion rate.
Thus, the outward diffusion of Fe must be driven by turbulent gas motions.

\bigskip

\subsubsection{General Constraints on Turbulence}

The diffusion coefficient can be written as $D=c_1 u l$
(where $c_1$ is a constant of order unity, $l$ is length scale of the largest 
eddies and $u$ is turbulent velocity for eddies of size $l$).  
Assuming the turbulence can be characterized by a local mixing length

\bigskip

\begin{inlinefigure}
\center{\includegraphics*[width=0.90\linewidth,bb=10 142 570 700,clip]{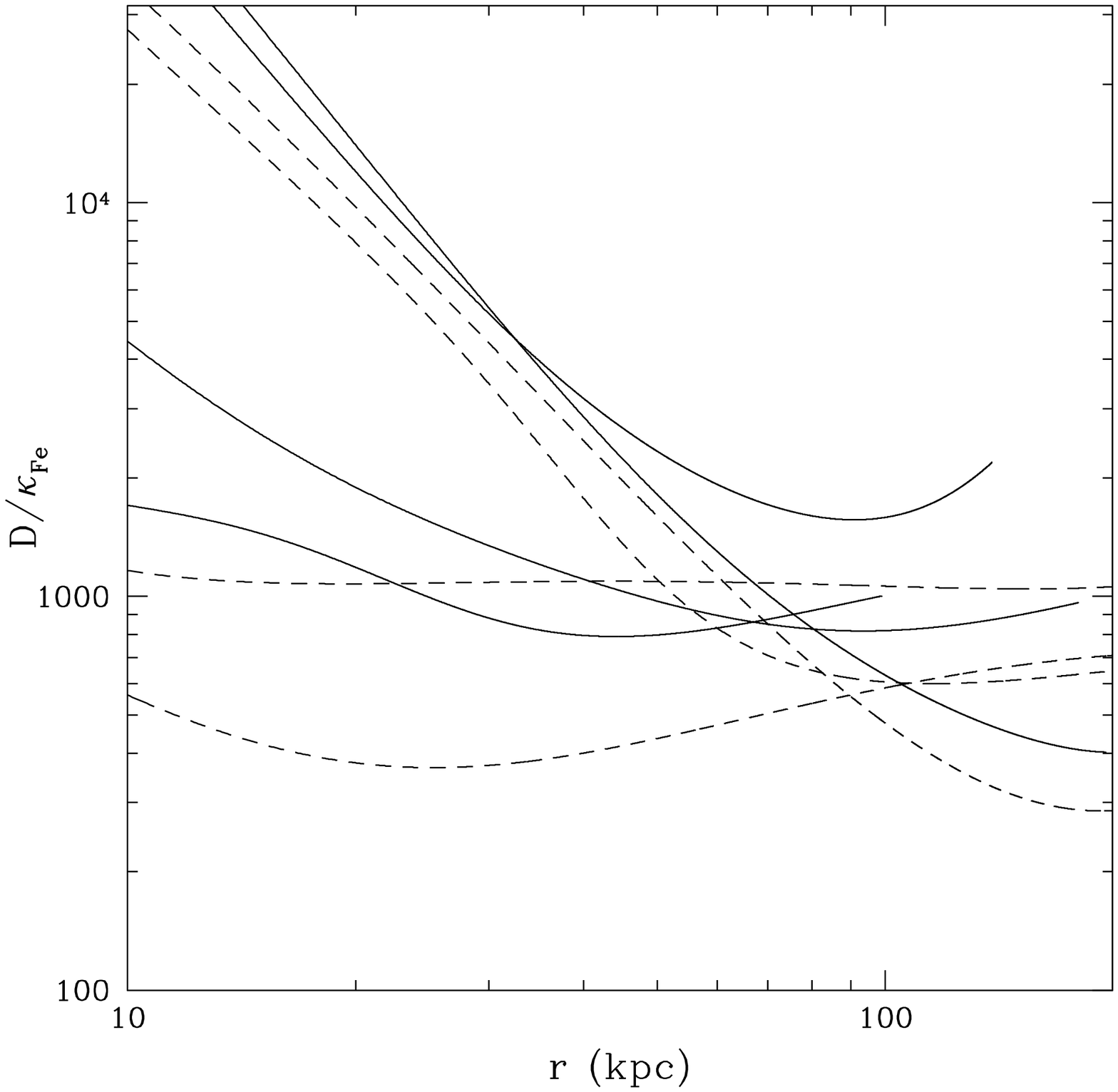}}
\caption{The ratio of the diffusion coefficient, $D$, to particle diffusion coefficient
for Fe, $\kappa_{Fe}=l_{Fe}v_{Fe}$,
where $l_{Fe}$ is the Fe-proton mean free path and $v_{Fe}$ is the thermal velocity
of the Fe ions. This figure shows that the diffusion of Fe is dominated by turbulent gas
motions ($D/\kappa_{Fe}$>1) at all radii in each cluster.
The line types are described in the caption to Fig. 1.}
\end{inlinefigure}

\begin{inlinefigure}
\center{\includegraphics*[width=0.90\linewidth,bb=10 142 570 700,clip]{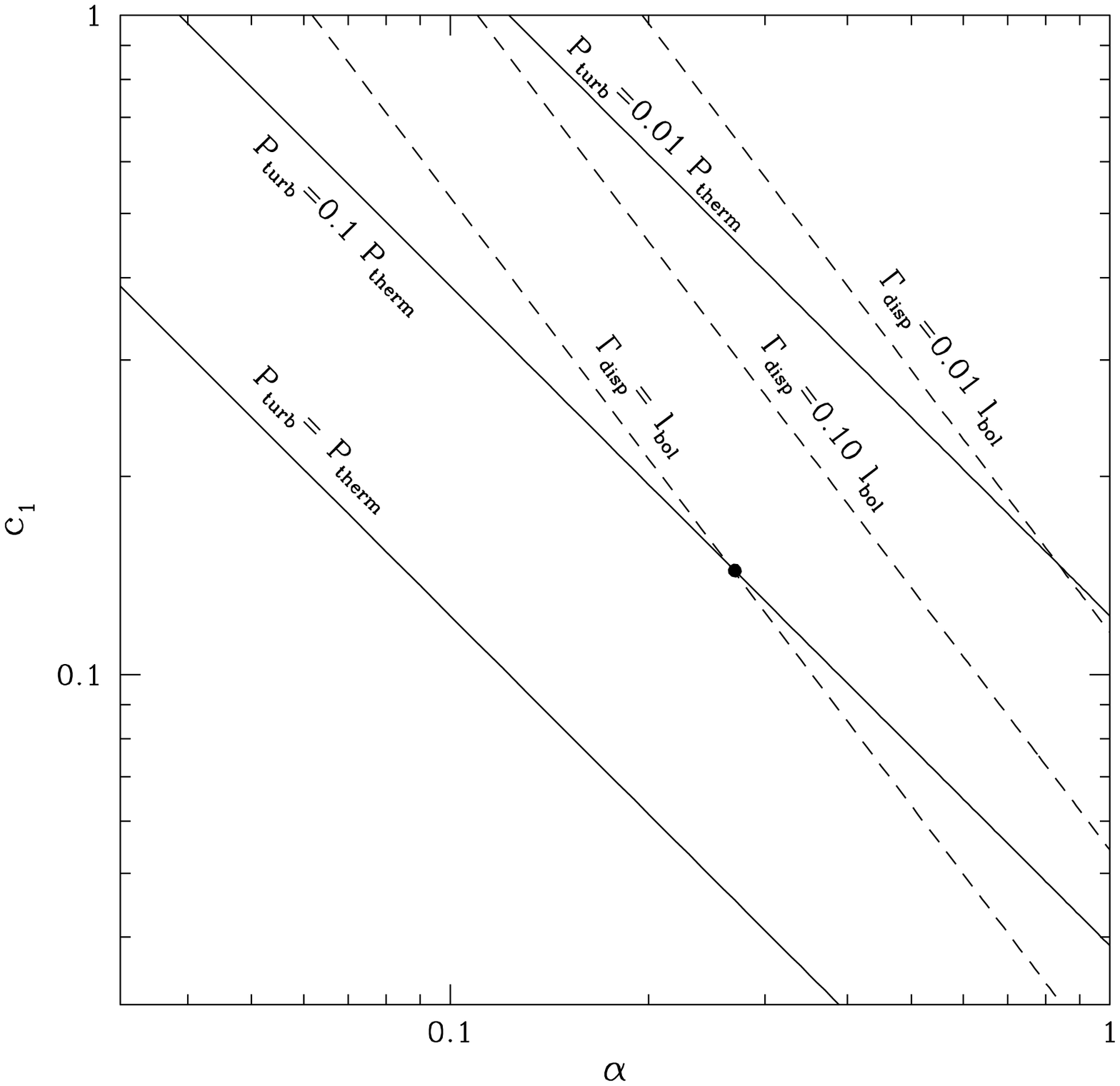}}
\caption{The solid lines indicate values of $\alpha$ and $c_1$ (defined in the text)
that produce constant values of $P_{turb}/P_{therm}$ using the conditions in A2199 at
a radius of 30~kpc.  The dashed lines correspond to constant values of $\Gamma_{disp}/l_{bol}$.
The solid point shows the values of $\alpha$ and $c_1$ required to completely balance
radiative cooling with a turbulent pressure equal to 10\% of the thermal pressure at a
radius of 30~kpc in A2199.}
\end{inlinefigure}

\bigskip

\noindent
prescription (see the discussion in Chandran 2005),
the size of the largest eddies can be written as $l=\alpha r$, where $\alpha$ is a parameter
between 0 and 1 and $r$ is the radial distance from the cluster center.
The turbulent eddies will cascade to smaller and smaller scales until 
the eddy turn over time is comparable to the viscous loss time and the turbulent
kinetic energy is dissipated at a rate per unit volume given by:

\begin{equation}
\Gamma_{diss} = c_2 \rho_g {{u^3} \over {l}} 
\end{equation}

\noindent
where $c_2$ is a constant of order unity. This equation can be written as: 

\begin{equation}
\Gamma_{diss} = \left( {{c_2} \over {c_1^3 \alpha^4}} \right) {{\rho_g D^3} \over {r^4}}
\end{equation}

\noindent
Written in this form, $\Gamma_{diss}$ is much more sensitive
to $c_1$ and $\alpha$ compared to $c_2$ and we therefore set $c_2=0.42$ based 
on the discussion in Dennis \& Chandran (2005).  

We can place some general 
constraints on $c_1$ and $\alpha$ based on the inferred turbulent gas pressure
and energy dissipation rate.  Based on recent work by Mahdavi et al. (2007)
and Churazov et al. (2007), gravitating masses within the cores
of rich clusters computed from X-ray 
data assuming only thermal pressure support are typically within 10\% of the 
masses derived from optical data and weak lensing.
This indicates that the sum of turbulent gas pressure, non-thermal gas pressure 
and magnetic pressure must be less than about 10\% of the thermal gas 
pressure in cluster cores.  The turbulent gas pressure can be written as:

\begin{equation}
P_{turb} = \left( {{1} \over {c_1 \alpha}} \right)^2 {{\rho D^2} \over {3 r^2}}
\end{equation}

\noindent
Using the gas properties in A2199 at a radius of 30~kpc and the diffusion coefficient
required to maintain a steady-state Fe

\begin{inlinefigure}
\center{\includegraphics*[width=0.90\linewidth,bb=10 142 570 700,clip]{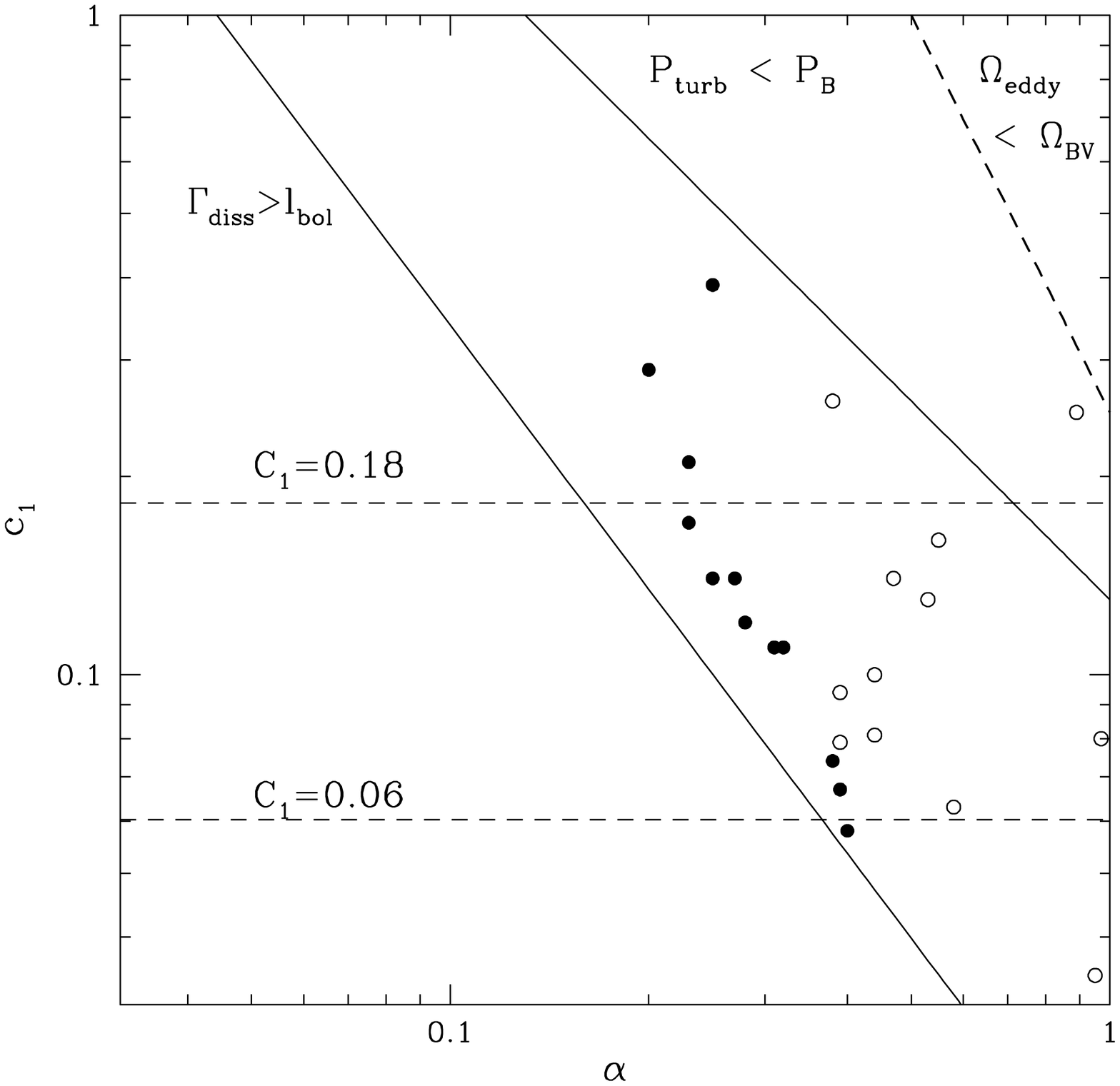}}
\caption{The solid points give the values of $\alpha$ and $c_1$ (defined in the text)
required to balance radiative cooling with heating by turbulent dissipation with a
turbulent pressure equal to 10\% of the thermal pressure at radii of 30, 50 and 70~kpc
in clusters with short cooling times.  The open points are the corresponding points for
clusters with long cooling times. The horizontal dashed lines bracket possible values
of $c_1$ from the literature.
Values of $\alpha$ and $c_1$ in the lower
left corner of the figure can be excluded since these values produce
heating by turbulent dissipation that exceeds radiative cooling at all radii
in all clusters.  Values of $\alpha$ and $c_1$ in the upper right hand corner of the
figure can also be excluded since these values produce turbulent pressures less than the
magnetic pressure (assuming $P_B=0.01 P_{therm}$). Values of $\alpha$ and $c_1$
in the extreme upper right hand corner of the figure yield eddy frequencies less
than the Brunt-Vaisala frequency.}
\end{inlinefigure}

\bigskip

\noindent
abundance profile, we plot 
lines of constant $P_{turb}/P_{therm}$ in the $c_1$ and $\alpha$ parameter space
in Fig. 12.  Also shown in Fig. 12 are lines of constant $\Gamma_{diss}/l_{bol}$,
where $l_{bol}$ is the X-ray bolometric luminosity per unit volume at the same 
location in A2199.  This figure shows that there is only a narrow strip in 
$\alpha$ and $c_1$ parameter space where the dissipation of turbulent gas motions 
is energetically important (i.e.,~ $0.1~l_{bol} < \Gamma_{disp} < l_{bol} $) and 
dynamically unimportant (i.e.,~ $P_{turb} < 0.1~P_{therm}$).  
This figure also shows that the dissipation of small turbulent eddies 
($\alpha  \lax 0.1$) can only 
balance radiative losses if $c_1$ is of order unity, while the dissipation of large 
eddies can balance radiative losses with a much broader range of $c_1$.

We have computed the values of $\alpha$ and $c_1$ required to balance radiative 
cooling through the dissipation of turbulence with $P_{turb}=0.1~P_{therm}$
at radii of 30, 50 and 70~kpc in each cluster.
The resulting data points are shown in Fig. 13.  This 
figure shows that most
of the clusters require eddies with $\alpha = 0.2-0.5$ and that there is a
trend that clusters with shorter cooling times require smaller eddies to suppress
radiative cooling. There are certain regions in the $c_1$ and $\alpha$ parameter space
that can be excluded. The region in the lower left corner of Fig. 13 indicates
regions where the dissipation of turbulence with $P_{turb}=0.1~P_{therm}$
exceeds the bolometric X-ray luminosity in all clusters at all radii.
The region in the extreme upper right of Fig. 13 can be excluded due to the 
effects of buoyancy.  Buoyancy suppresses turbulent diffusion if the Brunt-Vaisala
frequency, given by:

\begin{equation}
\omega_{BV}^2 = { {g} \over {\gamma} } {{d(lnP/\rho^{\gamma})} \over {dr}}
\end{equation}

\begin{inlinefigure}
\center{\includegraphics*[width=0.90\linewidth,bb=10 142 570 700,clip]{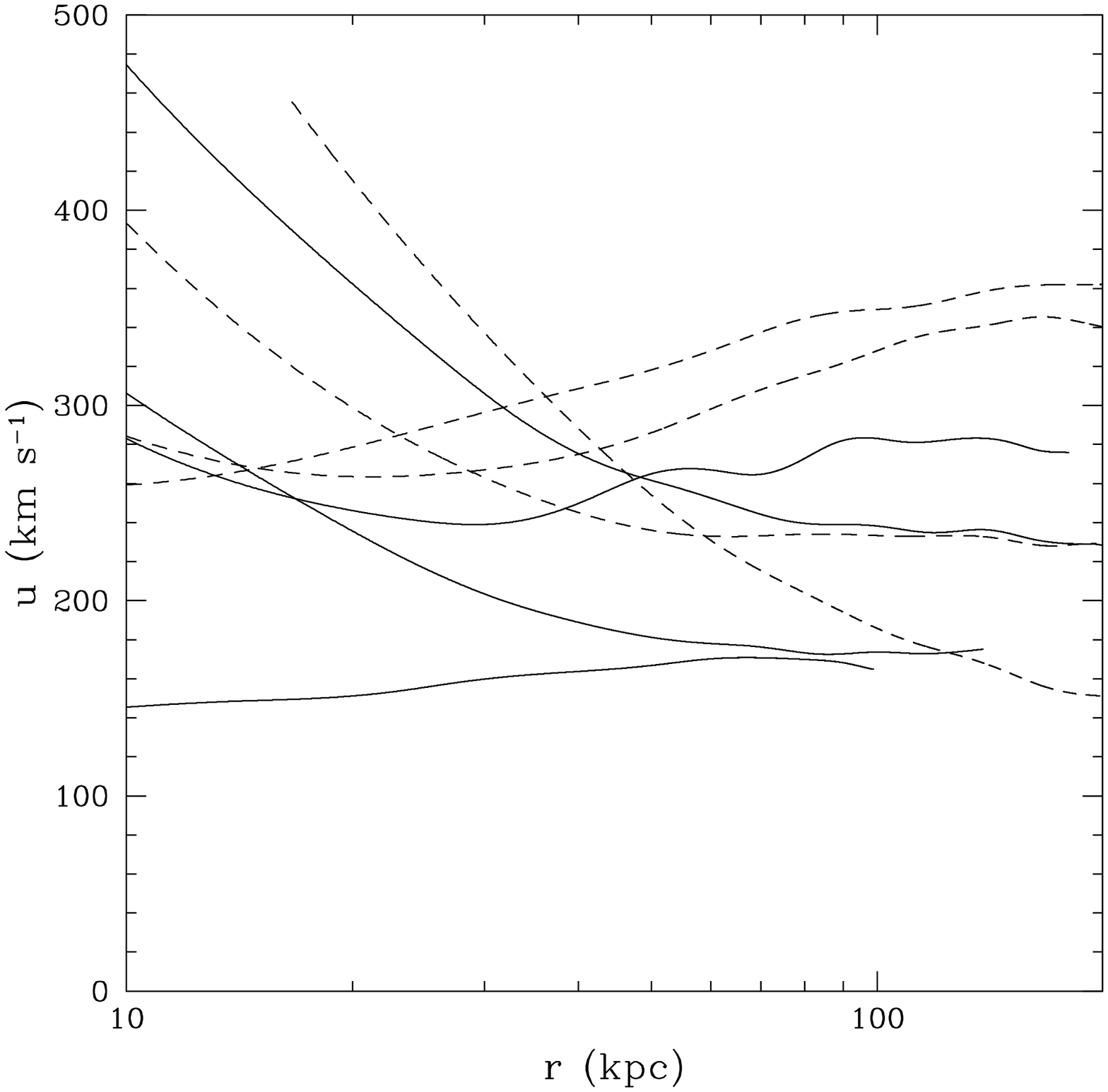}}
\caption{The turbulent velocity, $u$, required to balance radiative cooling with
heating by turbulent dissipation for each cluster in our sample.
The line types are described in the caption to Fig. 1.}
\end{inlinefigure}

\bigskip

\noindent
where $\gamma$ is the ratio of specific heats and $g$ is the acceleration
of gravity, is greater than the eddy turnover frequency, 
$\omega_{eddy}=u/l = D/(c_1 \alpha^2 r^2)$.
The region to the right of the heavy dashed line in Fig. 13 can be excluded
since these values of $\alpha$ and $c_1$ produce eddy turnover frequencies
less than the Brunt-Vaisala frequency in all clusters at all radii.
%%Dennis \& Chandran (2005) estimated that, even if $\omega_{eddy}/\omega_{BV}=1$, 
%%turbulent diffusion is only suppresed by 15\%.  
Turbulent diffusion is also suppressed when $P_{turb} < P_B$, where 
$P_B$ is the magnetic pressure.  Assuming $P_B=0.01~P_{therm}$ (which roughly 
corresponds to a magnetic field strength of $B=3-5~\mu G$ in the center of these clusters), 
a larger region in the upper right corner of the Fig. 13 can be excluded.
These calculations show that the suppression of turbulent 
diffusion by buoyancy is only significant for very low values of $B$.
Based on a review of the literature by Dennis \& Chandran (2005), they
find a range of estimates on $c_1$ from 0.06 to 0.18.  These values
are shown at horizontal dashed lines in Fig. 13.  Our calculations show that 
the region in the $\alpha$ and $c_1$ parameter space required for turbulent 
diffusion and dissipation to play a key role in clusters of galaxies
is mostly consistent with all of these constraints.

\subsubsection{Balancing Cooling Through Turbulent Dissipation}

In this section, we derive the properties of the turbulence 
required to locally balance radiative losses at all radii 
in each cluster using the diffusion coefficient necessary to maintain
a steady-state Fe abundance profile.
Substituting $l=D/c_1 u$ into eq. (7) gives $\Gamma_{diss}=c_1 c_2 \rho_g u^4 / D$.  
By setting $\Gamma_{diss}= f l_{bol}$ with $c_1=0.11$, $c_2=0.42$ and $f=1$,
the turbulent velocity required to balance radiative cooling
can be determined (see Fig. 14).  This figure shows that turbulent gas 
motions with $u =150 - 300$~km~s$^{-1}$ are required to locally balance 
radiative cooling in these clusters.  Even though 
this calculation assumes that turbulent dissipation completely balances 
radiative losses,
the estimated turbulent velocities are very insensitive to the fraction
of the cooling balanced by dissipation since 
$u \sim (f l_{bol})^{1/4}$. For example, if turbulent dissipation only 
balances 10\% of the radiative losses, then the resulting 
velocities would only be reduced by a factor of 2 below the values shown in Fig. 14.

\begin{inlinefigure}
\center{\includegraphics*[width=0.90\linewidth,bb=10 142 570 700,clip]{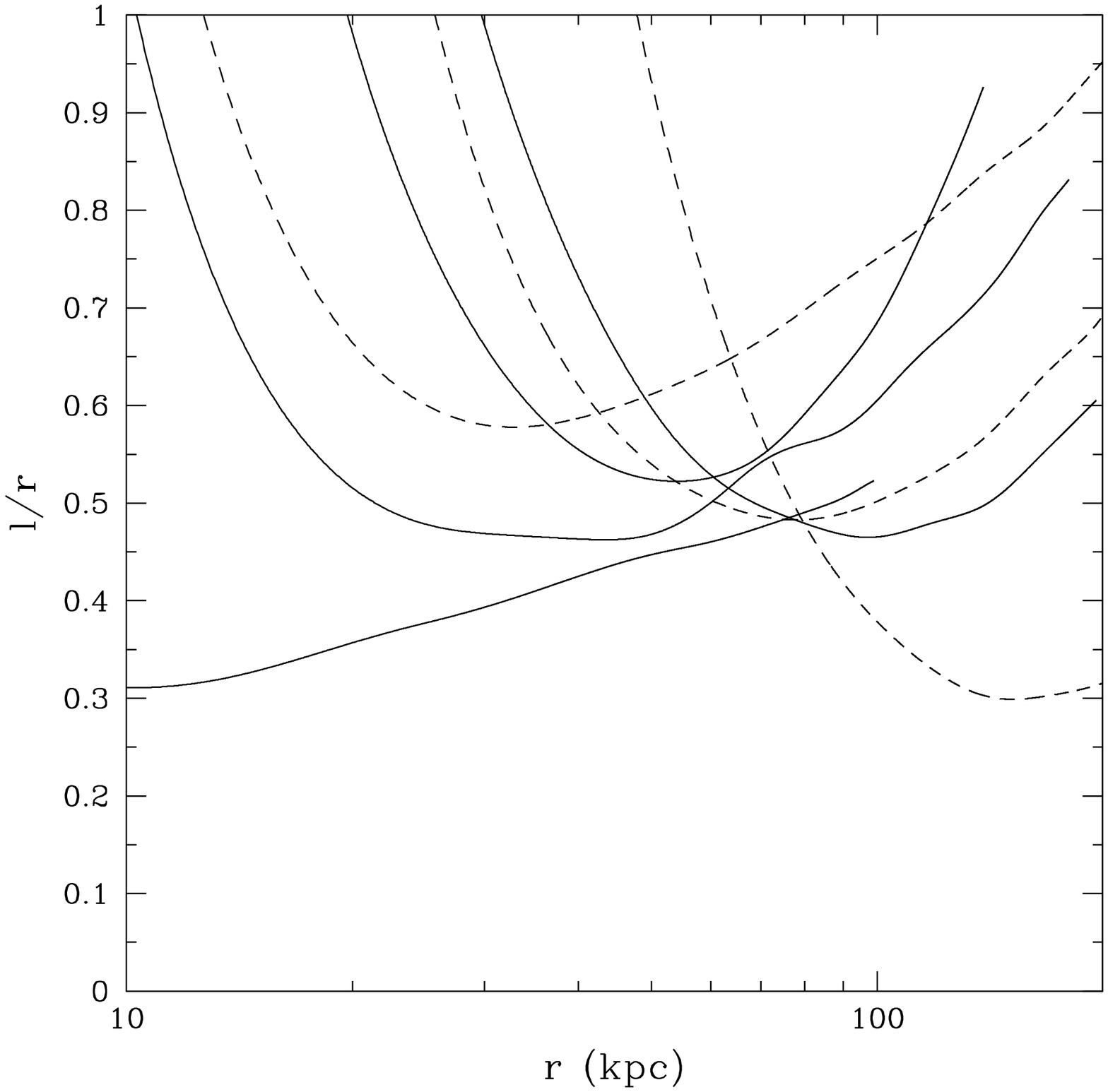}}
\caption{The ratio of the length scale of the largest eddies to the
cluster radius as a function of cluster radius for each cluster.
The line types are described in the caption to Fig. 1.}
\end{inlinefigure}

\bigskip

We can also estimate the scale length of the largest eddies from
$l=D/(c_1 u)$, which can be written as:

\begin{equation}
l = \left( { {D} \over {c_1} } \right)^{3/4} \left( { {c_2 \rho_g} \over {f l_{bol} } } \right)^{1/4}
\end{equation}

\noindent
Using the same values for $c_1$ and $c_2$ as given above, and $f=1$,
we have computed the ratio $l/r$ as a function of radius for each cluster 
(see Fig. 15).  It is apparent from this figure that large eddies 
with $\alpha > 0.5$ are required to locally balance radiative losses 
and maintain a steady-state Fe abundance profile.  There 
are regions
in Fig. 15 with $l/r > 1$, which are clearly unphysical.  Eq. (10) shows that 
the estimated eddy size is the most sensitive to $c_1$ and is fairly
insensitive to $c_2$ and $f$.  Increasing $c_1$ to 
0.18 (within the possible range noted by Dennis \& Chandran 2005)
would yield $l/r < 1$ in all clusters at all radii.  
The ratio of the turbulent pressure to the thermal pressure is shown 
in Fig. 16 using the turbulent velocities shown in Fig. 14.
This figure shows that $P_{turb}/P_{therm} < 0.1$ at most
radii, indicating that turbulent gas pressure is of little consequence in
estimating the gravitating masses of clusters even in cases when
turbulent dissipation completely compensates for radiative cooling.

\subsubsection{Balancing Cooling Through Turbulent Diffusion of Entropy}

Since the entropy increases outward in all of the clusters in our sample
(i.e., the gas is convectively stable), the outward diffusion of Fe will 
lead to an inward diffusion of entropy producing a heating rate per unit
volume given by:

\begin{equation}
\Gamma_{diff} = \vec{\nabla} \cdot (D \rho_g T \vec{\nabla} S)
\end{equation}

\noindent
Using the diffusion coefficient required to maintain a steady-state
Fe abundance profile, the ratio $\Gamma_{diff} / l_{bol} $ is shown as a 
function of radius for each cluster in Fig. 17.  In some regions 
(e.g., the central region in A496), the divergence term in eq. (11) is negative 
and turbulent diffusion locally cools the gas.  
Fig. 17 shows that the inward diffusion of entropy 
is an important heating mechanism in cluster cores.  In half of the 
clusters

\begin{inlinefigure}
\center{\includegraphics*[width=0.90\linewidth,bb=10 142 570 700,clip]{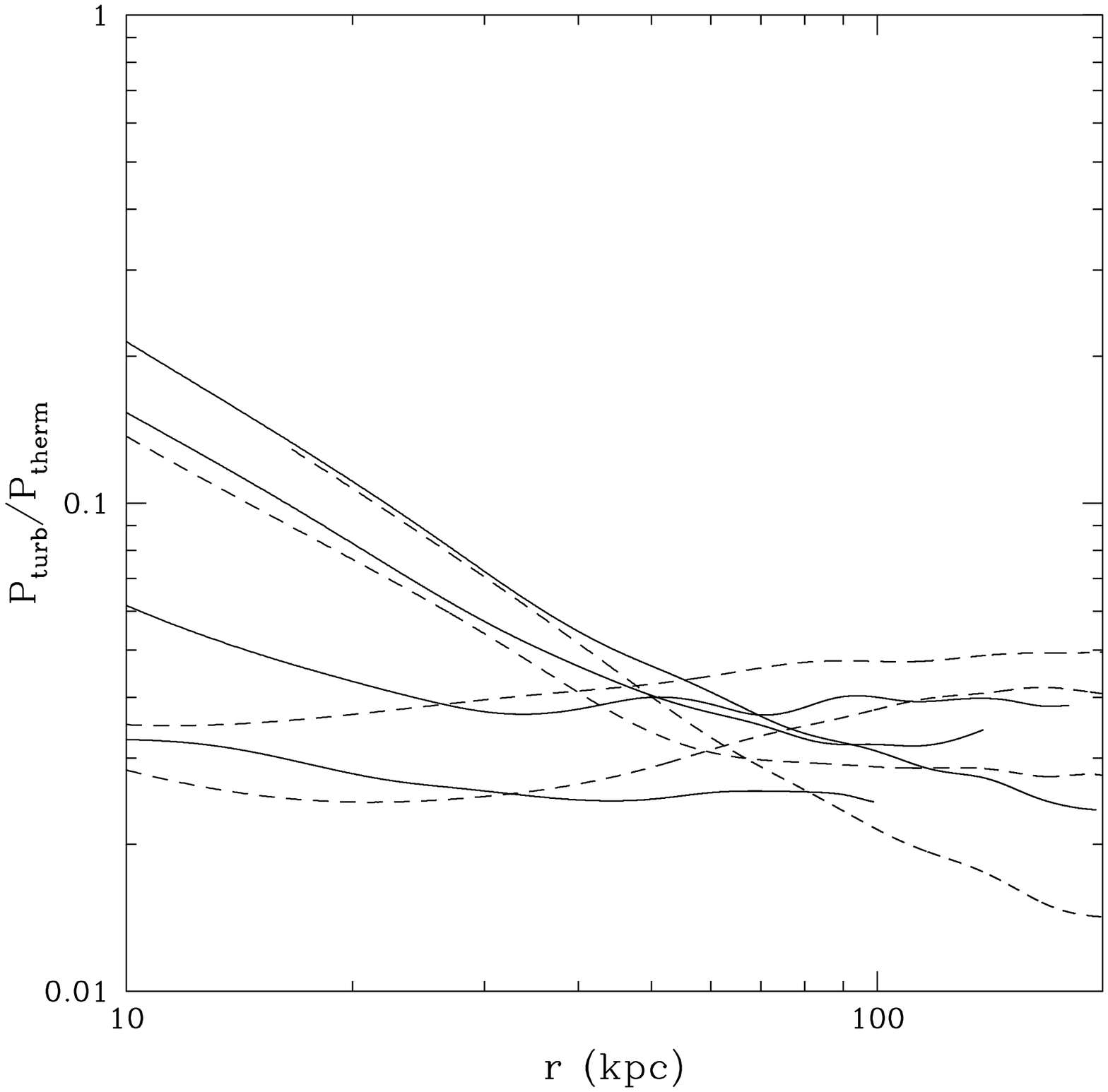}}
\caption{The ratio of turbulent gas pressure using the velocities in Fig. 14 to the
thermal gas pressure.  The line types are described in the caption to Fig. 1.}
\end{inlinefigure}

\bigskip

\noindent
with long cooling times (A3558 and A3571), turbulent diffusion
is sufficient to balance radiative losses at all radii and no additional
heating from turbulent dissipation is required.  In all other clusters,
some additional heating by turbulent dissipation is required.  The
required values of $\Gamma_{diss} / l_{bol}$ are shown in Fig. 17 in regions
where $\Gamma_{diff} / l_{bol} < 1 $.
The turbulent velocities necessary to balance cooling 
through a combination of turbulent dissipation and diffusion are shown in Fig. 18.  
These velocities are not significantly different than those shown in Fig. 14 which 
assumed that turbulent dissipation alone balances radiative losses.
In A496 and A3581, which have central regions with $\Gamma_{diff} < 0$ 
(see Fig. 17), larger velocities are required to balance cooling compared to the 
velocities shown in Fig. 14.

\section{Discussion and Conclusions}

Based on a systematic analysis of 8 clusters of galaxies, we find that
the excess Fe in the centers of these clusters is significantly 
more extended than the stellar distribution in the CDG.  There is a slight
trend that the 
central Fe excess is more extended in clusters with the long cooling times
compared to clusters with short cooling times.
The most likely origin for the excess Fe in cluster cores is enrichment
by SNe Ia from the CDG.  
Using the SNe Ia rate in early-type galaxies from 
Cappellaro et al. (1999), the total excess Fe mass within the central 100~kpc in 
these clusters can be produced by SNe Ia in 3-7~Gyr.   
We have not included the effects of ram pressure stripping 
of host galaxies in our calculations.  Based on a series of numerical simulations 
of clusters, 
Schindler et al. (2005) found that ram pressure stripping of host galaxies 
only increases the central Fe abundance by 0.07-0.10 (relative to the solar value).  
Most of the clusters in our sample have central Fe abundances approaching the solar 
value, so ram pressure stripping should not be a significant factor.
Domainko et al. (2006) also performed numerical simulations of the effects of ram 
pressure stripping on the enrichment of the intracluster medium in three cluster models
and found that the metallicity of the gas within the central 1.3~Mpc was only 
enriched by 10\%.  The ratio of total Si to

\begin{inlinefigure}
\center{\includegraphics*[width=0.90\linewidth,bb=10 142 570 700,clip]{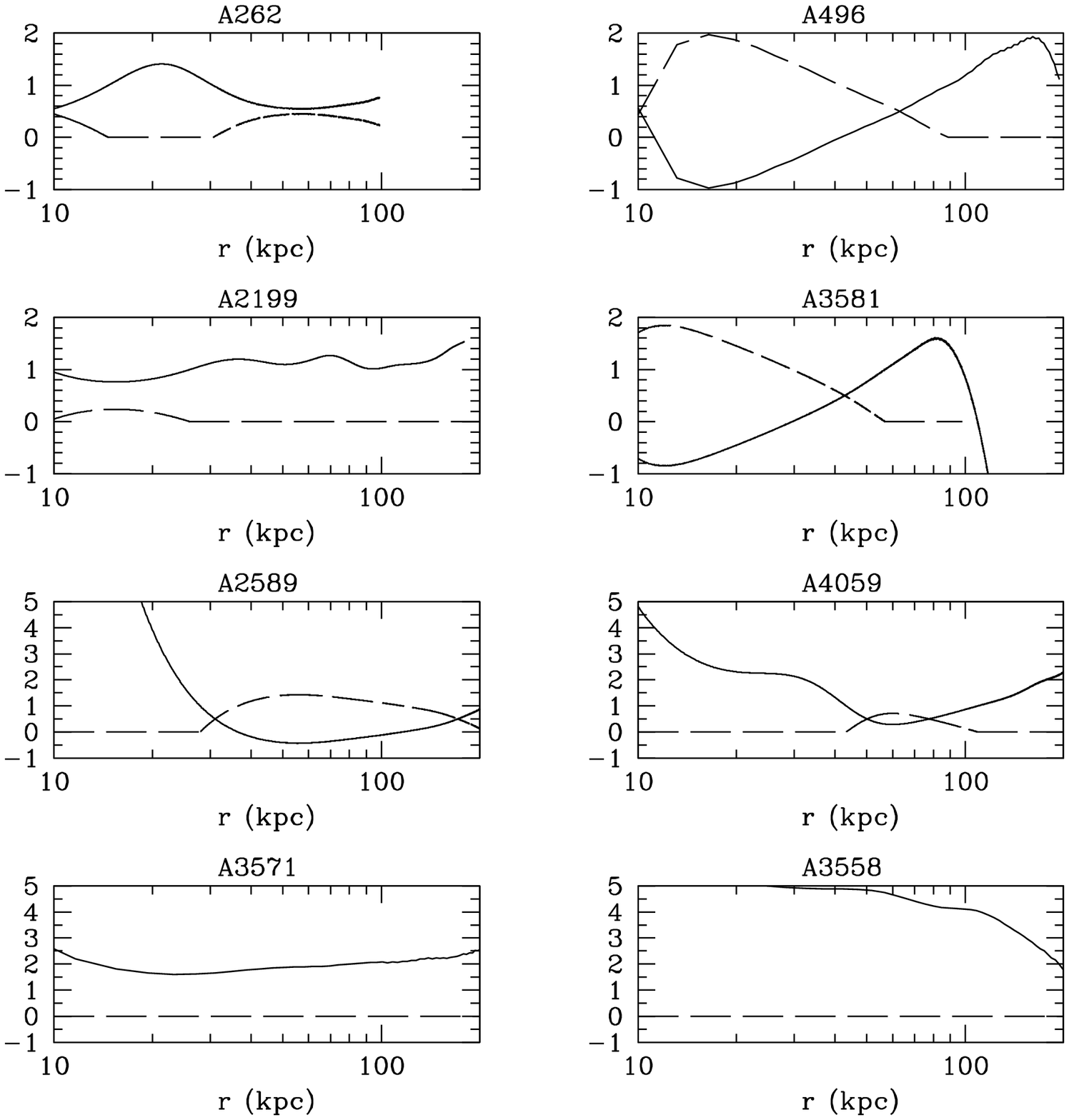}}
\caption{The ratio of the heating rate by turbulent diffusion of entropy per unit
volume to the bolometric X-ray per unit volume ($\Gamma_{diff}/l_{bol}$ - solid line)
as a function of radius for each cluster in our sample.  Also shown is the
ratio of the heating rate by turbulent dissipation per unit volume to the bolometric X-ray
luminosity per unit volume ($\Gamma_{diss}/l_{bol}$ - dashed line) in regions where
turbulent diffusion alone is insufficient to locally balance cooling.
The top 4 clusters have have central cooling times less than 1~Gyr
while the bottom 4 clusters have longer cooling times.}
\end{inlinefigure}

\bigskip

\noindent
Fe mass within the central 100~kpc in these 
clusters gives an average SNe Ia fraction of $f_{SNe Ia}=0.53$,
which is approximately twice the solar ratio.  Using the SNe Ia rate in 
Cappellaro et al. and $f_{SNe Ia}=0.53$, the total (Type Ia and core collapse)
supernova heating rate is, at most, 3-8\% of the bolometric X-ray luminosity 
within the central 100~kpc in these clusters.

In the absence of any heating mechanism, the excess Fe in the central regions of 
clusters should have the same distribution as the stars in the CDG, however
there is ample evidence from 

\bigskip

\begin{inlinefigure}
\center{\includegraphics*[width=0.90\linewidth,bb=10 142 570 700,clip]{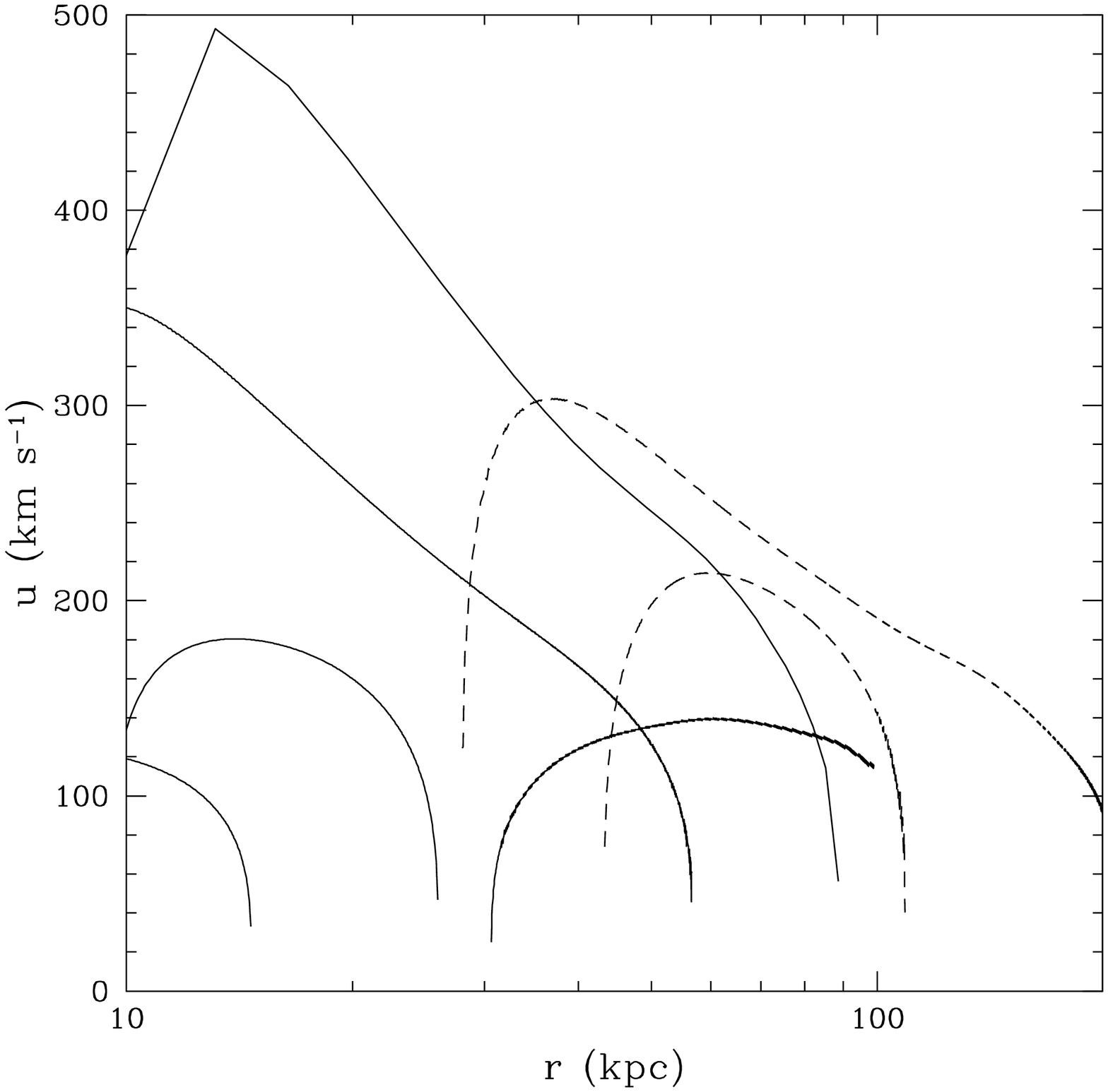}}
\caption{The turbulent velocity, $u$, required to balance radiative cooling with
a combination of heating by turbulent diffusion of entropy and dissipation
for all of the clusters in our sample.
The line types are described in the caption to Fig. 1.}
\end{inlinefigure}

\noindent
Chandra and XMM-Newton observations that 
AGN mechanical heating has a significant impact on the energetics of the 
gas in the central regions of clusters. Estimates on the total 
mechanical energy deposited into the gas from past AGN outbursts can be
obtained from the presently observed Fe distribution.
We find that the excess Fe must have undergone significant inflation
since being injected by SNe Ia, with $(r_f-r_i) / r_i \approx 1$ within 
the central 30-50~kpc, to reproduce the present Fe distribution in these clusters.
Using the observed gas temperature and density distributions,
along with the assumption of hydrostatic equilibrium, we determined
the gravitating mass distribution in each cluster and the 
potential energy difference between the present and initial excess Fe distributions.
Assuming that the Fe and gas expand together, we find that a total energy 
(thermal plus potential) of $10^{60} - 10^{61}$~erg must have been deposited
into the central 100~kpc in these clusters.  Since the required enrichment 
time for the excess Fe is approximately 5~Gyr, this gives an average mechanical power 
of $10^{43} - 10^{44}$~erg~s$^{-1}$ over this time.
This power is comparable to the ``cavity power" computed by Birzan et al. (2004) and 
Rafferty et al. (2006) for a sample of cool-core clusters.

The greater extent of the excess Fe in clusters compared to the stars
in the CDG could also arise from turbulent diffusion without a significant 
expansion of the total gas mass.  
Chandra and XMM-Newton observations have shown that AGNs in the centers
of clusters have a dramatic impact on the morphology and energetics of
the hot gas (see the review by McNamara \& Nulsen 2007).  Nuclear outbursts
can heat the gas through hydrodynamic shocks, the dissipation of 
sound waves and through the inflation and subsequent evolution
of buoyant bubbles.  The energy required to inflate a bubble is simply
the PdV work required to inflate the bubble plus the internal energy of
the material within the bubble, i.e., the enthalpy of the bubble.
Churazov et al. (2002) showed that the bubble enthalpy is transferred
to the hot gas via drag as the bubble buoyantly rises outward.
This energy initially takes the form of turbulent gas kinetic energy in the 
wake of the rising bubble.  The exact evolution of buoyant bubbles depends
on the whether the bubbles are filled with hot gas or relativistic particles
(Sijacki et al. 2008).  Heating by the dissipation of turbulent kinetic 
energy in clusters has been studied by 
Cho et al. (2003), Kim \& Narayan (2003), Voigt \& Fabian (2004),
Dennis \& Chandran (2005), Chandran (2005), but only Rebusco et al. (2005) 
and (2006) used the observed Fe distribution in clusters to constrain the properties
of the turbulence.  Since all of the clusters in our sample have a positive
entropy gradient, the outward diffusion of Fe will also lead to an inward diffusion
of entropy and further heating of the central gas.

We have obtained some general constraints on turbulence in cluster cores
by noting that the turbulent gas pressure must be less than 10\% of
the thermal gas pressure (to give agreement between cluster 
masses derived from X-ray data assuming only thermal pressure support 
and optical and weak lensing mass measurements), the turbulent gas 
pressure must be greater than the magnetic pressure (to prevent 
significant suppression of turbulent diffusion) and the eddy turn over 
frequency must be greater than the Brunt-Vaisala frequency.
These considerations show that there is only a narrow strip in $c_1$
and $\alpha$ parameter space (where $D=c_1 u l$ and $l=\alpha r$)
where turbulence can play a significant role in the gas energetics
in clusters without violating one of these constraints.  
We also find that small eddies (small values of $\alpha$) can only be 
energetically important if $c_1$ is close to unity, while larger eddies
must have smaller values of $c_1$ to be energetically important
without also being dynamically important.

We find that a diffusion coefficient of $10^{29} - 10^{30}$~cm$^2$~s$^{-1}$ 
is required to maintain a steady-state Fe abundance profile 
(i.e., the outward mass flux of Fe across a given surface
is equal to the mass injection rate of Fe from SNe Ia within that surface)
in our sample of 8 clusters.  These values are comparable with those derived
by Rebusco et al. (2005) and (2006) based on a comparison between a set of 
time-dependent models for the Fe enrichment in clusters and the presently 
observed Fe distributions.  The particle diffusion coefficient of Fe is 
approximately $\kappa_{Fe} \approx 1/3 \lambda_{Fe}v_{Fe}$, where $\lambda_{Fe}$ is 
the Fe-proton mean free path and $v_{Fe}$ is the thermal Fe velocity.  
In these clusters,
the particle diffusion coefficient of Fe is typically $10^{-3}$ times 
the diffusion coefficient estimated assuming a steady-state Fe abundance 
profile.  This shows that particle diffusion by itself is incapable 
of maintaining a steady-state Fe abundance profile and that the outward diffusion
of Fe must be driven by turbulence.  The dissipation of turbulent kinetic 
energy can be an important heating mechanism in the centers of clusters.
If the characteristic velocity of the turbulence is approximately 
150 - 300~km~s$^{-1}$, then the dissipation of turbulent kinetic 
energy can completely balance radiative cooling in these clusters.
Rebusco et al. (2005) and (2006) derived similar turbulent velocities for their
sample of 8 groups and clusters.  Even if the velocity of the turbulence is 
sufficient to completely balance radiative cooling, we find that the resulting 
turbulent gas pressure is less than 10\% of the thermal gas pressure at most 
radii in these clusters.

In addition to heating by turbulent dissipation, we find that heating
by turbulent diffusion of entropy is also an important heating 
mechanism in cluster cores.
Radiative cooling can be completely balanced by turbulent diffusion alone
in 2 out of the 4 clusters in our sample with central cooling times longer that 1~Gyr.
All of the other clusters in our sample require
some additional heating by turbulent dissipation to fully balance cooling.
Dennis \& Chandran (2005) calculated the significance of turbulent
diffusion, turbulent dissipation and heat conduction in three
cool-core clusters assuming the turbulence in clusters is generated by 
cosmic rays produced by the central supermassive black hole that mix
with the thermal gas.  They parameterized the 
length scale of the largest eddies as $l = l_0 + \alpha r$ and generated
a set of models with different values of $\alpha$ for each cluster.
In their models with $\alpha > 0.5$, they found that heating by turbulent 
diffusion is the dominant heating mechanism, while in models with $\alpha < 0.5$, 
heating by turbulent dissipation dominants. This is easily understood
since heating by turbulent diffusion is directly proportional to $D$, and
hence $l$ (see eq. 11), while heating by turbulent dissipation is inversely proportional 
to $l$ (see eq. 6), with all other parameters held fixed.  
We directly calculate $l$ at each radius in each cluster by assuming that the 
Fe abundance profiles are in steady-state and that radiative losses are balanced
through a combination of turbulent dissipation and diffusion.  
In general, we find $\alpha > 0.5$ at most radii in most clusters which increases
the significance of heating by turbulent diffusion.

In the above discussion, we assumed that clusters are in steady-state.
While this is a reasonable assumption for examining clusters on 
timescales longer that the time between AGN outbursts, it is possible 
that the level of turbulence in clusters is periodic and peaks shortly after 
an AGN outburst.  Deep Chandra observations of several clusters have detected
a series of X-ray cavities emanating from the central AGN. 
If turbulence is driven by the dredging up
of low entropy, enriched gas behind a buoyant AGN inflated bubble
(along with the subsequent infall of higher entropy, less enriched gas),
then the initial heating triggered by an AGN outburst is due to the turbulent 
diffusion of entropy.  Hydrodynamic isotropic Kolmogorov turbulence 
cascades according to $u \sim l^{1/3}$.
Heating by turbulent diffusion scales as $\Gamma_{diff} \sim u l \sim l^{4/3}$,
while heating by turbulent dissipation scales as $\Gamma_{diff} \sim u^3/l \sim l^0$.
Thus, heating by turbulent diffusion decreases as the turbulence cascades
to smaller scales while heating by turbulent dissipation remains
a constant.  However, both the break up of turbulent eddies to smaller scales
and the dissipation of turbulence occur on approximately an eddy-turn over time, 
so both processes essentially operate together.

\bigskip

This work was supported in part by NASA grant GO7-8127X. PEJN also acknowledges NASA 
grant NAS8-01130. The authors would also like to thank the anonymous referee for
many useful suggestions and comments.

\newpage

\appendix
\section*{Change in Thermal Energy of the Gas During Bulk Expansion}

Assume that the gas initially fills a volume $V_i$ with a radius of
$R_i$.  After a period of bulk expansion, the gas fills a final
volume of $V_f$ with a radius of $R_f$.  The difference in thermal energy
between the final and initial states is simply:

\begin{equation}
\Delta U = {{3} \over {2}} \left[ \int_0^{V_f} P_f(r_f) dV - \int_0^{V_i} P_i(r_i) dV \right]
\end{equation}

\noindent
where $P_i$ is the initial pressure of the gas and $P_f$ is the final
gas pressure.  Assuming the gas is in hydrostatic equilibrium
before and after the bulk expansion, then the pressure is given by:

\begin{equation}
P(r) = \int_{r}^{\infty} \rho(r)g(r)dr.
\end{equation}

\noindent
where $\rho(r)$ is the gas density and $g(r)$ is the acceleration of gravity.
Using eq. (2) in the first integral on the right hand side of eq. (2) gives:

\begin{equation}
\int_0^{V_f} P_f(r_f) dV = \int_0^{V_f} dV \int_{r_f}^{\infty} \rho_f(r)g(r)dr.
\end{equation}

\noindent
Integration by parts gives:

\begin{equation}
\int_0^{V_f} P_f(r_f) dV = \left[ V \int_{r_f}^{\infty} \rho_f(r)g(r)dr \right]_0^{V_f} 
+ \int_0^{V_f} V(r_f) \rho_f(r_f)g(r_f)dr_f
\end{equation}

\noindent
which becomes:

\begin{equation}
\int_0^{V_f} P_f(r) dV = {{4 \pi} \over {3}} \left[ R_f^3 \int_{R_f}^{\infty} \rho_f(r)g(r)dr
+ \int_0^{R_f} r_f^3 \rho_f(r_f)g(r_f)dr_f \right].
\end{equation}

\begin{equation}
= V_f P_f(R_f) + {{4 \pi} \over {3}} \int_0^{R_f} r_f^3 \rho_f(r_f)g(r_f)dr_f.
\end{equation}

\noindent
Using $dM_f = 4 \pi \rho_f(r_f)r_f^2 dr_f$ and $v_k^2(r_f) = g(r_f)r_f$,
where $v_k$ is the Keplerian velocity, eq. (5) becomes:

\begin{equation}
\int_0^{V_f} P_f(r) dV  = V_f P_f(R_f) + {{1} \over {3}} \int_0^{M_f} v_k^2(r_f)dM_f.
\end{equation}

\noindent
Performing the same calculation for the second term on the right hand side of
eq. (1) and taking the difference gives:

\begin{equation}
\Delta U = {{3} \over {2}}  \left[ V_f P_f(R_f) - V_i P_i(R_i) \right] 
+ {{1} \over {2}} \left[\int_0^{M_f} v^2_k(r_f)dM_f - \int_0^{M_i} v^2_k(r_i)dM_i \right].
\end{equation}

\noindent
Since $M_f = M_i$, this becomes:

\begin{equation}
\Delta U =  {{3} \over {2}}  \left[ V_f P_f(R_f) - V_i P_i(R_i) \right]
+ {{1} \over {2}} \int_0^{M_f} (v_k^2(r_f) - v_k^2(r_i))dM_f.
\end{equation}

\noindent
At large radii, $r_f-r_i \rightarrow 0$, and the first term converges
to zero.

\clearpage

\end{document}